\documentclass[journal]{IEEEtran}
\usepackage{amsmath}
\usepackage{amsfonts}
\usepackage{amssymb}
\usepackage{graphicx}
\usepackage{picins}
\usepackage[ruled, vlined]{algorithm2e}
\usepackage{url}
\usepackage{color}

\newtheorem{defi}{\noindent \textbf{Definition}}
\newtheorem{prop}{\noindent \textbf{Proposition}}

\begin{document}

\title{SAFA: a Semi-Asynchronous Protocol for Fast Federated Learning with Low Overhead}
\author{Wentai~Wu, \IEEEmembership{Student Member,~IEEE}, Ligang~He, \IEEEmembership{Member,~IEEE}, Weiwei~Lin, Rui~Mao, Carsten Maple, and Stephen~Jarvis, \IEEEmembership{Member,~IEEE}
\thanks{W. Wu, L. He (corresponding author) and S. Jarvis are with the Department of Computer Science, the University of Warwick. W. Lin is with the School of Computer Science and Technology at the South China University of Technology. R. Mao is with the College of Computer Science and Software Engineering, Shenzhen University. Carsten Maple is with Warwick Manufacturer Group at the University of Warwick.}}

\markboth{This is a preprint version of the work \href{https://www.computer.org/csdl/journal/tc/5555/01/09093123/1jNu0qlnwSk}{10.1109/TC.2020.2994391} published in the IEEE TC by \copyright IEEE}%
{}

\maketitle

\begin{abstract}
Federated learning (FL) has attracted increasing attention as a promising approach to driving a vast number of end devices with artificial intelligence. However, it is very challenging to guarantee the efficiency of FL considering the unreliable nature of end devices while the cost of device-server communication cannot be neglected. In this paper, we propose SAFA, a semi-asynchronous FL protocol, to address the problems in federated learning such as low round efficiency and poor convergence rate in extreme conditions (e.g., clients dropping offline frequently). We introduce novel designs in the steps of model distribution, client selection and global aggregation to mitigate the impacts of stragglers, crashes and model staleness in order to boost efficiency and improve the quality of the global model. We have conducted extensive experiments with typical machine learning tasks. The results demonstrate that the proposed protocol is effective in terms of shortening federated round duration, reducing local resource wastage, and improving the accuracy of the global model at an acceptable communication cost.
\end{abstract}


\section{Introduction}
With the prevalence of Internet of Things (IoT), the advance in Machine Learning (ML) techniques stimulates the demand of compute capacity significantly from a broad range of applications which more or less integrate Artificial Intelligent (AI) into the edge and end devices to empower their underlying business logic. By 2022, more than 80\% of enterprise IoT projects are expected to have AI components embedded \cite{Gartner}. Also, it has been an emerging trend that users are becoming more sensitive to the data privacy protection mechanism of AI applications, while their performance, in many cases, is still expected to be guaranteed in the first place. 

It is promising for intelligent applications to learn their models on massively distributed data. However, there are still several obstacles to date. First, it is unrealistic to collect decentralized data constantly from all the end devices and store them in a centralized location, which can probably cause a lot of potential risks (e.g., data leakage) and poses privacy threats to end users. Second, it could be communication-intensive to train a global model using traditional optimization methods, no matter in a cloud-central or a distributed manner. On-cloud centralized training incurs heavy load (as well as big risks) in data transfer when moving the data from the edge of network to the cloud, whilst most distributed optimization approaches incur fairly frequent communications between devices and the cloud in order to exchange gradients (of a mini-batch, typically) and weights. However, in practical circumstances (e.g., edge computing environments), the devices are hardly reliable and the cost of communication can be prohibitive. For example, devices may drop offline intermittently and data transfer is charged in cellular networks.

Privacy concerns may prohibit moving data outside local devices. Machine learning in such environments is challenging due to the following properties: 1) \textbf{Unbalanced and biased data distribution}: the end devices may own a variable amount of on-device data and the distribution of the data in different devices may be different; 2) \textbf{Massive distribution}: it is usual to see a huge fleet of disparate devices at the edge as participants; 3) \textbf{Unreliability}: either the devices themselves and the connection to the cloud are unreliable. End devices could opt out occasionally or go offline unexpectedly. The communication could be expensive.

Federated Learning (FL) \cite{FL_comm}\cite{origin_FL}, a promising framework, was proposed by Google to address the aforementioned challenges. The work presents a distributed solution (i.e., Federated Optimization) to optimizing a global machine learning model without moving data out of local devices, and introduces \textit{FedAvg} as an optimization protocol in federated setting. Rather than collecting gradients from clients (i.e., end devices), FedAvg adopts a different approach, in which multiple iterations of local updates (using gradient descent) are followed by a global aggregation that takes a weighted average of the resulting models from the clients. An obvious advantage of FedAvg is the reduction of communication frequency. FedAvg and many implementations of FL systems (e.g., \cite{FL_system}) adopt synchronous training protocols to avoid the prohibitive number of updates. Although synchronous protocols seem to be the natural choice for the FL setting, a number of limitations stand out as follows: 1) \textbf{Unreliable fraction of effective participants}: in each round, the server selects a fraction of clients randomly to perform local training and expects them to commit their local training results. However, the number of clients which manage to commit their results are very uncertain given the unreliable nature of end devices; 2) \textbf{Low round efficiency}: To aggregate the local results at the end of each round, FedAvg has to wait for all selected clients to finish, among which there may be stragglers while the crashed ones may never respond. Consequently, the global learning progress is suspended until a timeout threshold is reached; 3) \textbf{Under-utilization of clients}: With random selection, many capable clients are likely to remain idle even if they are ready and willing to participate in the training; 4) \textbf{Progress waste}: The selected clients may not finish local training in time, and the progress made could be wasted because the client will be forced to overwrite its local model with the global model next time when the client is selected again.

In this paper, we propose a Semi-Asynchronous Federated Averaging (SAFA) protocol based on FedAvg \cite{origin_FL} to achieve fast, lag-tolerant federated optimization. SAFA takes advantage of several efficiency-boosting features from asynchronous machine learning approaches (e.g., \cite{async_multi_task}\cite{FedAsync}) while making use of a refined pace steering mechanism to mitigate the impact of straggling clients and stale models (i.e., staleness \cite{FedAsync}) on the global learning progress. Moreover, we adopt a novel aggregation algorithm that exploits a cache structure (in the cloud) to bypass a fraction of client updates so as to improve convergence rate at a low cost of communication. The main contributions of our work are outlined as follows:
\begin{itemize}
\item We take into account the unreliability and heterogeneity of end devices and propose a Semi-Asynchronous Federated Averaging (SAFA) protocol to alleviate the staleness, boost efficiency and better utilize the progress made by stragglers.
\item We introduce a simple hyper-parameter, \textit{lag tolerance}, to flexibly control the behavior of SAFA protocol. We also empirically analyze the impact of lag tolerance on SAFA by observing how it affects the critical metrics such as synchronization ratio and version variance.
\item We conducted extensive experiments to evaluate SAFA on several typical machine learning tasks in multiple FL settings varying from tiny to relatively large-scale edge environments. SAFA is evaluated in terms of several important metrics such as model accuracy, round efficiency and communication cost.
\end{itemize}

The rest of this paper is organized as follows: Section II summarizes some relevant studies on distributed learning and federated learning. In Section III, we formulate the optimization problem for FL, detail the design of SAFA and analyze the impact of the hyper-parameter in SAFA. In Section IV, we present and discuss the experimental results. We conclude this paper in section V.

\section{Related Work}
The fusion of Edge Computing and Artificial Intelligence (i.e., Edge Intelligence \cite{EI1}\cite{EI2}) has emerged as a new focus of research ever since we began to realize the potential benefits of sinking the computation to and outside the edge whilst the increasing capacity of end devices makes it natural to empower them with AI and support the applications such as intelligent surveillance \cite{surveil} and mobile keyboard prediction \cite{keyboard} at the edge.

Distributed machine learning is believed to be an ideal solution for big data analytics according the rule "moving computation closer to data". However, the majority of distributed ML approaches (e.g., \cite{distML1}\cite{distML2}\cite{DC-ASGD}) claim their efficacy based on the conditions such as homogeneity \cite{DC-ASGD}, high-performance nodes, ultra-fast connections \cite{QSGD} and so on, which are unrealistic in edge computing or IoT environment. In fact, end devices in an edge environment can be fairly unreliable, highly heterogeneous in performance and have limited communication. The limitation of data access is another prominent issue. Many distributed ML approaches cannot achieve the desired accuracy without making the entire dataset available to every worker. However, it is impossible in many situations to gather the data from a massive number of distributed devices given the expensive communication (via cellular networks) and, most importantly, the data privacy requirements by end users \cite{IBM_FL}.

Federated Learning (FL) \cite{origin_FL}, first proposed by Google, is a new approach to fitting machine learning into the edge. The survey by Zhou et al. \cite{EI3} summarizes recent studies on edge intelligence and lists FL as one of the most uprising technologies for distributed training at the edge. As the primitive FL protocol, \textit{FedAvg} \cite{origin_FL} was designed to perform synchronous optimization in federated settings. Xie et al. \cite{FedAsync} proposed \textit{FedAsync}, an asynchronous federated optimization scheme that regularizes local optimization and adopts the non-blocking update of the global model. A similar protocol has been exploited by Sprague et al. \cite{Async_geo} in a geo-spatial application for training a global model asynchronously, allowing the devices to join halfway. However, the main issue of the asynchronous approaches is that the server may receive too many local updates sent from a massive number of clients that remain active, which could overwhelm the server but with little benefit to the model convergence.  

In terms of model accuracy, Chen et al. \cite{exp_SSGD} experimentally demonstrated that synchronous Stochastic Gradient Descent (SGD) can outperform asynchronous approaches in the data center setting, which to some extent inspired the synchronous design of FL. A number of variants have been proposed to mitigate the deficiencies of FL from different aspects such as round efficiency \cite{FedCS} and communication cost \cite{FL_comm}. Wang et al. \cite{FL_WangS} proposed a control algorithm that adaptively determines the interval of global aggregation under a given resource budget. To address the inefficiency of FL under poor wireless channel conditions, Nishio and Yonetani \cite{FedCS} implemented a mobile edge computing (MEC) framework in which a protocol is designed to filter out slow clients based on the estimation of the clients' work time at the selection stage and consequently shorten round length. However, their scheme relies on accurate estimation and does not take the client unreliability into account.  

How to speedup the convergence rate of FL remains an open challenge. On this point, we argue that the optimization mechanisms for traditional distributed SGD have great potential in FL. For example, gradient staleness control has been shown critical to guarantee convergence \cite{staleness1}. Dutta et al. \cite{staleness2} theoretically characterized the trade-off between reducing error (by including more stragglers) and shortening run time (by bounding staleness). Wang et al. \cite{staleness3} refined ASGD by modulating the learning rate based on the staleness of incoming gradients. Smith et al. \cite{MOCHA} proposed MOCHA, a fault- and straggler-tolerant multi-task learning method without forging a global model. Chen et al. \cite{backup_workers} introduced backup workers to reduce server waiting time in synchronous stochastic optimization. Inspired by these approaches, we investigate the impact of straggling clients and model staleness on FL, and design a fast FL protocol that is well adapted to unreliable environments.

\section{The SAFA Protocol}
The proposed Semi-Asynchronous Federated Averaging (SAFA) protocol is designed to solve the global optimization problem as below:
\begin{equation}
\label{target_f} 
    \arg\min_{w \in \mathbb{R}^d} \frac{1}{n} \sum_{i=1}^n f(w; x_i, y_i)
\end{equation}
where $w$ denotes the parameters of the global model (the number of parameters = $d$), $f(w; x_i, y_i)$ represents the loss of the inference on sample $(x_i,y_i)$ made by the model with $w$ as its parameters. Note that data samples are distributed among disparate end devices, which are called clients in FL settings. Let $M$ denote the set of $m$ clients, and $D_j$ the partition of data residing in client $j$, then the target function can be rewritten as:
\begin{equation}
\label{target_f2}
    \arg\min_{w \in \mathbb{R}^d} \frac{1}{n} \sum_{j=1}^m \sum_{i \in D_j} f(w; x_i, y_i)
\end{equation}

Note that the problem definition here is in accordance with \cite{origin_FL}, but is different from \cite{FedAsync}. Xie et al. \cite{FedAsync} define their target function as the average of the average loss (on local partitions), which is fair at the local partition level but not the case at the sample level, because data samples in small local partitions take larger weights in their target function.

\begin{figure*}[ht]
    \centering
    \includegraphics[width=5.0in]{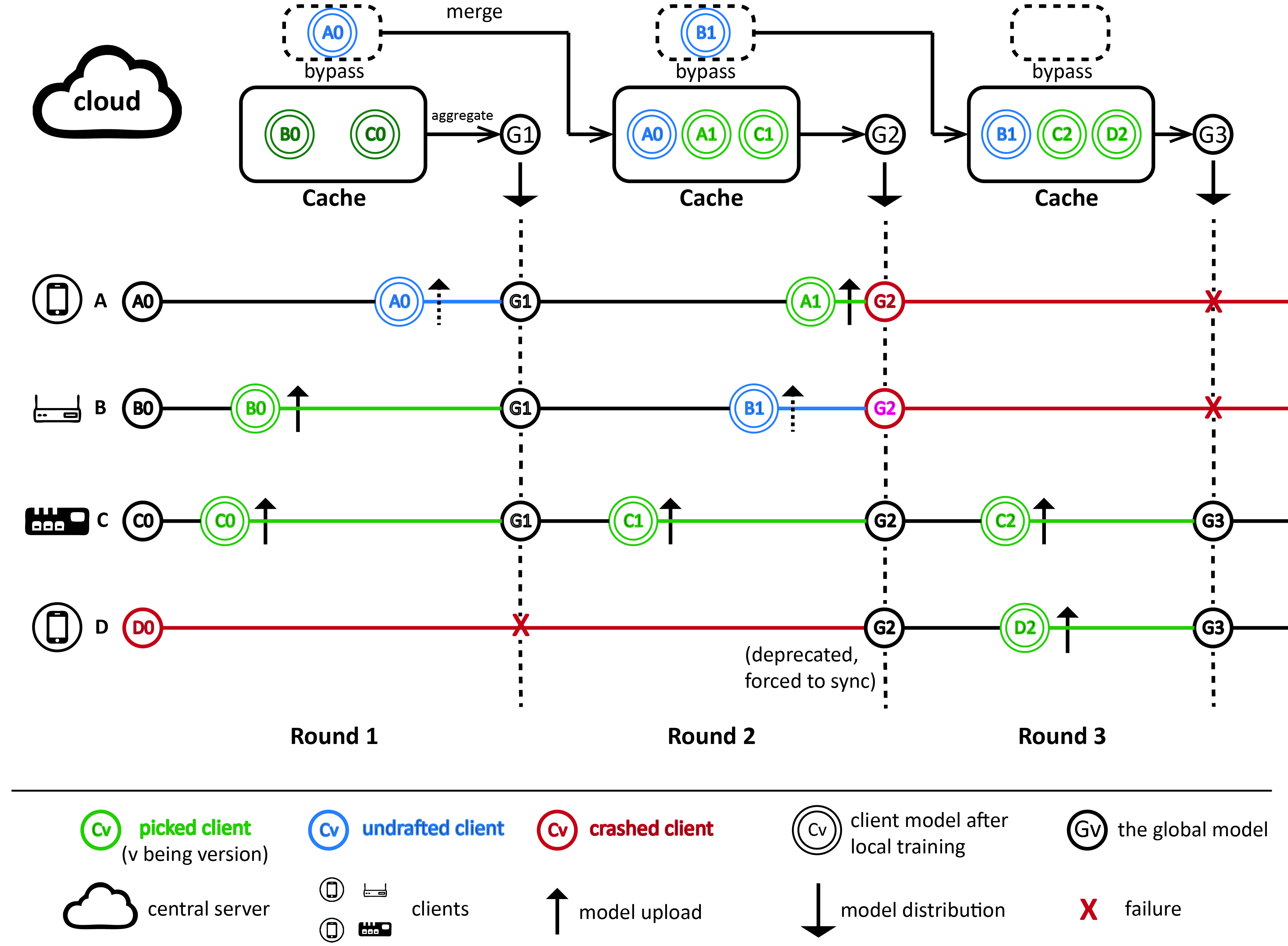}
    \caption{The diagram of SAFA protocol showing the interaction between the cloud and end clients in different states}
    \label{fig:SAFA_diagram}
\end{figure*}

In this section, we present the workflow of SAFA with the underlying design principles in detail. As a refined FL protocol, SAFA consists of three operations: lag-tolerant model distribution, post-training client selection and discriminative aggregation. A typical FL process driven by SAFA is shown in Fig. \ref{fig:SAFA_diagram}. We will use this diagram as an example throughout this paper to illustrate our FL training process.   

For clarity, we list in Table \ref{tab:symbols} the symbols frequently used in this paper.

\begin{table}[ht]
\centering
\caption{List of symbols}
\begin{tabular}{ l l } 
 \hline
 Symbol 	  	 & Description \\ 
 \hline
 $D$			 & the complete dataset \\
 $n$			 & the size of $D$ (i.e., $n=|D|$) \\
 $D_i$			 & the data partition on client $i$ \\
 $n_i$			 & the size of client $i$'s local partition\\
 $M$			 & the set of clients (i.e., end devices) \\
 $m$			 & total number of clients \\
 $v_i$			 & the version of client $i$'s local model \\
 $M_v$			 & the set of clients whose model version is $v$ \\
 $P$			 & the set of picked clients \\
 $P_v$			 & the set of picked clients of version $v$ \\
 $K$			 & the set of crashed clients \\
 $K_v$			 & the set of crashed clients of version $v$ \\
 $W$ 			 & the set of clients that complete local training \\
 $Q$			 & the set of undrafted clients \\
 $Q_v$			 & the set of undrafted clients of version $v$ \\
 $w$			 & parameters of the global model \\
 $w_k$			 & parameters of the local model on client $k$ \\
 \hline
\end{tabular}
\label{tab:symbols}
\end{table}

\subsection{Lag-tolerant Model Distribution}
In the federated setting, an important guarantee for the convergence of the global model is the quality of local models. The quality of a local model depends on whether local training is sufficient (i.e., training sufficiency) and on the starting model on which local training is based. Training sufficiency can be achieved by allowing adequate local iterations (i.e., epochs), while the version control is a non-trivial task. The original FL algorithm \cite{origin_FL} prevents outdated clients with stale models from committing, which simplifies FL process but also throttles the potential of accelerating convergence.

Motivated by the problem, we first present a lag-tolerant model distribution algorithm which does not always enforce synchronization (i.e., allows some clients to stay asynchronous with the cloud) and is tolerant to outdated local models (i.e., staleness). The key idea is to develop a better way to get the stragglers (i.e., clients with stale models) involved in the model aggregation and leverage their progress for faster federated learning. In this paper, we refer \textit{stragglers} to the clients who are slow and still conducting local training based on an outdated model. Normally, the clients are supposed to start epochs of training based on the latest global model received from the server. However, device crashes or network problems generate the stragglers inevitably. 

With a version-based criterion, SAFA only requires specific clients to retrieve the latest global model from the server. Before a round of local training starts, the server classifies all clients into three states (or categories) based on their current versions: \textit{Up-to-date}, \textit{deprecated} and \textit{tolerable}, which are defined as follows. 

\defi \textit{Up-to-date clients}: the clients that have completed the previous round of local training (and submitted models successfully) are reckoned up-to-date at the start of this round.

\defi \textit{Deprecated clients}: the clients that still base local training on the models that are too stale compared to the version of the global model.

\defi \textit{tolerable clients}: the clients that do not base local training on the latest global model, but the model version they are based on is not too old either. This is a state that stands between \textit{Up-to-date} and \textit{Deprecated}. 

SAFA only requires the up-to-date and deprecated clients to synchronize with the server, while the tolerable clients stay asynchronous with the server. This is why SAFA is called a semi-asynchronous distributed training scheme. We let up-to-date clients synchronize with the server in order to prevent model divergence \cite{origin_FL}. Deprecated clients are forced to synchronize so that the global model will not be poisoned by the seriously outdated local models. 

After a round of local training is completed on device, the clients will then be labeled \textit{picked}, \textit{undrafted} or \textit{crashed} based on the result of client selection in SAFA, which is a post-training process. The server tags clients with these labels after the selection quota is met or the round time limit is reached. The picked clients are those whose local training results in this round are selected to be used in the following aggregation step. The undrafted clients are those whose local training results are not selected but still get cached by the server for future use. Crashed clients are those who fail to finish a round of local training - clients can either opt out or drop offline intermittently (i.e., any time during training) with a certain probability (which we refer to as crash probability).

In Fig. \ref{fig:SAFA_diagram} we illustrate the workflow of SAFA with four end devices: clients A to D, with which the system is to perform several federated rounds of training. Clients start local training from their local model versioned A0 to D0 (the initial model version for each round of local training is depicted in the figure by a single circle with the model version in the middle). After a client completes its local training, local parameters are updated (depicted by double circles with the model version in it) and are uploaded to the server (depicted by upwards arrows). The server selects submitted results only from a portion of clients (the client portion is 50\% in this example) to update the global model. The clients whose updates are selected are tagged as picked clients (colored green), for example, clients B and C in the 1st round. The selected updates are placed in a cache structure by the server. The cache maintains the entries of the latest local models uploaded from the picked clients and will be used for aggregation. The clients whose results are not selected are undrafted clients (colored blue), e.g., client A in round 1 and client B in round 2. Updates from these clients are stored in the bypass structure to avoid futile work locally. The clients who cannot complete their local training due to any reason (such as opt-out or network failure) are crashed clients (highlighted red), such as client D in the 1st round. 

Each round ends with a new version of global model (i.e., G1 to G3 in this diagram), which, at the start of the next round, will be distributed to (i.e., synchronized with) the up-to-date and deprecated clients. In the first round, for example, A, B and C successfully complete local training and upload their updates (in spite of A being undrafted), thus they become up-to-date clients (i.e., tagged \textit{up-to-date} by the server). The results of undrafted clients will not be merged into the global model in the upcoming aggregation step, but may take effect in future rounds via a bypass structure (squares with dashed lines) that saves these updates temporarily. The bypass will merge with the cache right after the current aggregation step before the next round starts.

In this example, we assume the maximum tolerable version lag is 2. In Fig. \ref{fig:SAFA_diagram}, client D does not manage to finish local training in two rounds. Therefore, it is tagged \textit{deprecated} and forced to synchronize with the server, which means client D needs to replace its stale local model with the latest global model. To decide whether a local update should be accepted, here we adopt a simple criterion based on the difference between the versions of the global model and the local model, which is called \textit{lag tolerance}. Therefore, the deprecated clients are those whose local version lags behind the version of the global model by more than the specified \textit{lag tolerance}. Specifically, our lag-tolerant distribution principle can be formulated as follows:

\begin{equation}
	w_k(t) =
	\begin{cases}
		w(t-1) 		& \text{if } k \in \bigcup_{v=t-1}M_v, \text{or } k \in \bigcup_{v<t-\tau}M_v, \\
					&\text{ // up-to-date or deprecated clients} \\
		w_k(t-1) 	& \text{if } k \in \bigcup_{t-\tau\leq v<t-1}M_v \\ 
					&\text{ // tolerable clients}\\
    \end{cases}
\label{eq:lag-tol_dist}
\end{equation}
where $w(t-1)$ denotes the latest global model parameters (i.e., the aggregation result from last round) upon the start of round $t$, and $w_k$ denotes the parameters of client $k$'s local model; $\tau$ stands for \textit{lag tolerance}, which is the only hyper-parameter in SAFA. The lag-tolerant model distribution forces the up-to-date and deprecated clients to adopt the latest global model as the base model for the next round of training, while the tolerable clients can continue to work on their previous local results. The hyper-parameter \textit{lag tolerance} in some ways controls the tradeoff between communication overhead and the convergence rate of federated optimization. If it is set too small, the server may suffer heavy downlink transmission as the portion of deprecated clients increases. If it is set too large, the convergence of the global model could be unsteady. The impact of \textit{Lag tolerance} will be analyzed later with empirical studies.

\subsection{Client Selection}
An important property of end devices is unreliability, which means that they occasionally drop offline for some reasons such as power outage (or low battery level), inaccessible network or manual shutdown/opt-out of training. In this paper, we refer to these temporarily unavailable states as \textit{crashed}. Every client has a certain probability to crash in each round of training. For clients that stay active and connected to the central server (throughout a round of training), we assume they are always able to finish the task assigned within a certain period of time (otherwise they are also reckoned crashed). 

The population of committed updates should be carefully limited considering a huge fleet of end devices \cite{FL_big_survey}. McMahan et al. \cite{origin_FL} use a hyper-parameter $C$ to control the maximum fraction of clients allowed to participate in one round of training. Moreover, $C$ serves as the criterion in the \textit{FedAvg} \cite{origin_FL} protocol by which the server keeps waiting for selected clients to end an global round. In our approach, we retain this hyper-parameter but no longer apply it as a hard constraint. Instead, we release the restriction to allow all clients to participate if they are willing to, and enable the central server to end a round once $C$-fraction of updates have been received.

Apparently, the efficiency of federated optimization is closely associated to the fraction of picked clients. One may think that we can set $C$ to a large value (e.g., close to 1.0) and pick as many clients into each round as possible. However, it is neither realistic nor beneficial to do so. On the one hand, allowing more clients to participate increases the potential risk of uplink congestion and the communication cost as well. In each round, the server may have to wait for more clients among which some may never respond (because picked clients could crash midway). On the other hand, involving a large number of updates leads to limited benefit to the global model especially in the last few rounds before convergence \cite{origin_FL}.

It is notable that the fraction of selected clients (called selection fraction) is not equivalent to the actual fraction of clients that finish local training and commit their models in time. In an unreliable environment, picked clients can crash halfway in their training progress or fail to upload their trained models. In this paper, we define a metric termed Effective Update Ratio (EUR) to measure the fraction of effective updates from the local (i.e., all clients) to the cloud (i.e., central server(s)).

\begin{equation}
  EUR = \frac{|P - P \cap K|}{|M|}
  \label{eq:eur}
\end{equation}
where $P$ and $K$ are the sets of picked and crashed clients, respectively. Obviously $EUR$ is positively correlated with the size of $P$ and negatively correlated with that of $K$. As mentioned, simply increasing the pick fraction can bring about problems in the FL context, while the crash of clients is not predictable or controllable (improving client stability is beyond the scope of this paper). As a solution, we propose to let the central server collect local update after local training instead of randomly selecting clients at the very beginning of a global round. This means the server does not need to wait for those designated clients for aggregation but are able to execute the aggregation step once it has received a $C$-fraction of update. Our post-training selection effectively decouples the server with the selected clients and consequently improves $EUR$, which facilitates faster convergence of the federated optimization. Another advantage of doing so is a significant boost of round efficiency in the case the clients crash with a fairly high probability. Based on the outcome of selection (before the aggregation step is carried out), the server tags the clients with three different labels: \textit{crashed}, \textit{picked} and \textit{undrafted}. Only picked clients are eligible to update its corresponding cache entry right before the aggregation conducted by the central server. Undrafted clients also commit their updates but their updates will bypass the following aggregation.

\begin{figure}[ht]
    \centering
    \includegraphics[width=2.1in]{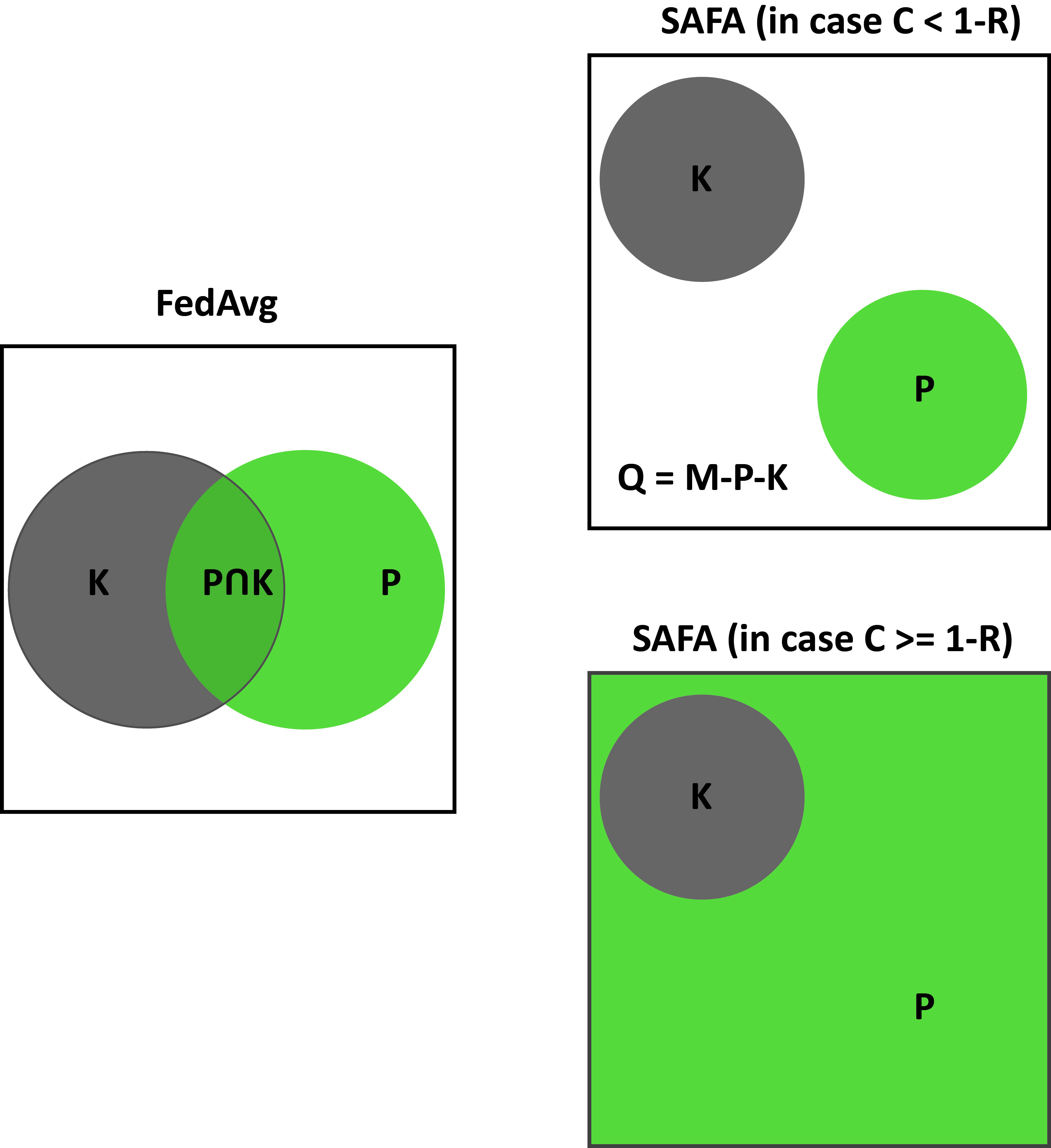}
    \caption{The diagram shows the client selection policies by FedAvg and SAFA. The selection fraction is $C$ in both policies. SAFA picks $C$-fraction of clients if no less than \textit{C} fraction of clients uploaded their local models (i.e., updates). Otherwise it will pick all clients that have committed the local models. A square in the diagram represents the full client set $M$. \textit{K} and \textit{P} are the sets of crashed and picked clients, respectively.}
    \label{fig:venn}
\end{figure}

Considering the "selection-ahead-of-training" scheme used in the synchronous FL (e.g., FedAvg \cite{origin_FL}), its effective update ratio, according to Eq. (\ref{eq:eur}), is $C(1-\frac{|K|}{|M|})$. By contrast, SAFA adopts a "selection-after-training" scheme that theoretically yields the value of $EUR$ as follows:

\begin{equation}
  EUR = 
  \begin{cases}
		1 - R			& \text{if } C \geq 1-R,\\
		C 				& \text{if } C < 1-R\\
    \end{cases}
  \label{eq:eur_safa}
\end{equation}
where $C$ is the selection fraction and $R$ denotes the crash ratio over all the clients (i.e., $R = \frac{|K|}{|M|}$). Fig. \ref{fig:venn} demonstrates how SAFA promotes the effective update ratio in FL - involving as many clients as possible to fulfill the fraction $C$.

From Fig. \ref{fig:venn} and considering (\ref{eq:eur}), we can see a clear improvement of $EUR$ by SAFA, which minimizes the negative impact of clients' failure. Nevertheless, extremely high crash ratio of clients will still cause a low value of $EUR$ even with our selection method. The phenomenon will be analyzed later in our experiment section.

As figured out by Bonawitz et al. \cite{FL_system}, bias is introduced if each device is equally likely to participate each round because they differ in performance and network access privilege. The problem remains if we merely use the client selection method mentioned above. Therefore, we further propose to alleviate the bias using a compensatory client selection algorithm. The principle is simple -- higher priority is given to those clients that are less involved. In each round the server maintains a list of IDs of clients that missed the previous round of training, and their updates will be picked prior to others for the coming aggregation. The pseudo-code of our selection policy is shown in Algorithm \ref{algo:compens_select}.

\begin{algorithm}[ht] 
\caption{Compensatory First-Come-First-Merge (CFCFM) client selection}
  \DontPrintSemicolon
  \SetKwInOut{Input}{Input}
  \SetKwInOut{Output}{Output}
  \SetAlgoLined
  \Input{round number $t$, client set $M$, last-round picked clients $P(t-1)$, selecting fraction $C$, round deadline $T_{lim}$}
  \Output{clients to pick $P(t)$}
  $P(t) = \varnothing$\;
  $Q(t) = \varnothing$\;
  $quota = C \cdot |M|$\;
  \While{$|P(t)| < quota \mathrm{~and~} T_{round} < T_{lim}$}{
    Await new updates\;
    $w'_k \leftarrow$ update arrives from client $k$\;
    \uIf{$k \mathrm{~not~in~} P(t-1)$}{
      add $k$ to $P(t)$\;
    }
    \Else{
      add $k$ to $Q(t)$\;
    }
  } 
  
  \If{$|P(t)|<quota$}{
    Sort $Q(t)$ by arrival time\;
    $q \leftarrow quota-|P(t)|$\;
    $P'(t) \leftarrow$ first $q$ clients in $Q(t)$\;
    $Q(t) \leftarrow Q(t) - P'(t)$ \;
    $P(t) \leftarrow P(t) + P'(t)$ \;
  }
  return $P(t)$
\label{algo:compens_select}
\end{algorithm}

In the selection, we stop involving more clients once the quota has been met, namely \textit{C}-fraction of clients have been selected from $P(t-1)\cap W(t)$. Otherwise the algorithm continues to wait and accept the updates (until a deadline is reached) from the rest of clients which, in practice, will arrive at the cloud successively. 

\subsection{Discriminative Aggregation}
After a round of local training completes, the server has received a collection of updates from the end devices. We adopt three steps to aggregate local updates. The first step is the pre-aggregation cache update, which overwrites the corresponding entries (for storing model parameters) of the selected clients in the cache. In the second step, the updates stored in the cache are aggregated. In the third step, the undrafted updates are placed in the cache, which can be used in the next round of global model aggregation. Since the picked and undrafted updates are treated in a different manner in the aggregation, it is called the three-step discriminative aggregation, which is formally formulated as follows:

\noindent (1) \textit{Pre-aggregation Cache Update}:
\begin{equation}
	w^*_k(t) =
	\begin{cases}
		w'_k(t) 		& \text{if } k \in P(t),\\
		w(t-1)			& \text{if } k \in \bigcup_{v<t-\tau}M_v(t),\\
		w^*_k(t-1)		& \textrm{otherwise}
    \end{cases}
  \label{eq:pre-aggre}
\end{equation}
where $w^*_k(t)$ denotes the $k$-th entry of the cache structure (see Fig. \ref{fig:SAFA_diagram}), and $w'_k(t)$ denotes the trained local model at round $t$. Entries of deprecated clients will be replaced with the global model $w(t-1)$.

\noindent (2) \textit{SAFA Aggregation}:
\begin{equation}
	w(t) = \sum_{k=1}^m \frac{n_k}{n} w^*_k(t)
  \label{eq:aggre}
\end{equation}

\noindent (3) \textit{Post-aggregation Cache Update}:
\begin{equation}
	w^*_k(t+1) =
	\begin{cases}
		w'_k(t) 		& \text{if } k \in Q(t),\\
		w^*_k(t)		& \textrm{otherwise}
    \end{cases}
  \label{eq:post-aggre}
\end{equation}
where $P(t)$, $Q(t)$, and $K(t)$ denote the sets of picked, undrafted, and crashed clients, respectively in round $t$. 

For SAFA, there are three cases of changes in the cache after a global around $t$. For picked clients, their updates will be kept in the cache after being merged into the global model. For undrafted clients, the updates will not take effect in this round but will be carried to the next round by the post-aggregation step. For the crashed clients, their entries stay unchanged only if they have not been deprecated. Otherwise these entries will be replaced by the global model (i.e., $w(t-1)$ in Eq. \ref{eq:pre-aggre}) to avoid heavy staleness.

\begin{algorithm}[ht] 
\caption{Semi-Asynchronous Federated Averaging (SAFA) protocol}
  \DontPrintSemicolon
  \SetKwInOut{Input}{Input}
  \SetKwInOut{Output}{Output}
  \SetAlgoLined
  \Input{maximum number of rounds $r$, client set $M$, local mini-batch size $B$, number of local epochs $E$, learning rate $\eta$, lag tolerance $\tau$}
  \Output{finalized global model}
  \smallskip
  
  \bf Server process:\qquad \rm// running on the central server\;
    \rm Initializes client connections\;
    \rm Initializes global model $w(0)$ and the cache\;
  	\For{\rm round $t=1$ to $r$}{
      Distributes $w(t-1)$ according to Eq. (\ref{eq:lag-tol_dist}) given $\tau$\;
      \For{\rm each client $k$ in $M$ \bf in parallel}{
      	$w'_k(t)=$ \bf{client\_update}$(k,w_k(t))$
      }
      Collects and selects client updates using CFCFM\;
      Updates cache according to Eq. (\ref{eq:pre-aggre})\;
      Performs aggregation and get $w(t)$ using Eq. (\ref{eq:aggre})\;
      Updates cache according to Eq. (\ref{eq:post-aggre})\;
  	} 
  	\rm return $w(r)$
  	
  \smallskip
  \bf Client process:\qquad \rm// running on the client $k$\;
  	\bf client\_update$(k,w_k)$:\;
  	$B_k \leftarrow$ \rm batches from $D_k$ of size $B$\;
  	\For{\rm epoch $e=1$ to $E$}{
    	\For{\rm batch $b$ in $B_k$}{
      	$w_k = w_k - \eta \nabla f(w_k; b)$\;
    	}
  	}
  	$w'_k = w_k$\;
  	\rm return $w'_k$ to the server
  
  \label{algo:SAFA}
\end{algorithm}

Now we can present the complete workflow of the proposed SAFA protocol outlined in Algorithm \ref{algo:SAFA}. The server orchestrates the process holistically in rounds. At the beginning of each round, the server first checks the version of clients and distributes the latest global model in a lag-tolerant manner (see Eq. \ref{eq:lag-tol_dist}) given the hyper-parameter $\tau$. Then the server begins to listen and collects the updates (i.e., trained local model) from clients. Clients train their native models on local datasets using the gradient descent method. Based on Algorithm \ref{algo:compens_select}, the clients missing the previous round will have the priority to be selected to meet the pre-set fraction $C$. Following the client selection, the server then executes the three-step discriminative aggregation, which merges all the entries in the cache into the global model, i.e., $w(t)$, and updates the cache entries of undrafted clients.

\subsection{Analysis of Lag tolerance}
We analyze the impact of \textit{lag tolerance} from different perspectives. As mentioned, this hyper-parameter is crucial to the pace-steering of the SAFA protocol. When \textit{lag tolerance} is small, clients/models become deprecated frequently, resulting in relatively high cost in model distribution. If it is set to a big value, the server will be very tolerant to the stragglers, which will probably cause high variance in the versions of local models and consequently slow down the convergence of the global model. Thus, we introduce two holistic metrics: Synchronization Ratio (SR) and Version Variance (VV). $SR$ measures the usage of downlink by which the global model is distributed to the edge of network. $VV$ is defined based on the version distribution of local updates. For SAFA, we formulate SR and VV as follows:

\begin{equation}
	SR_{SAFA} = 
	\frac{1}{rm} \sum_{t=1}^{r} 
	(|\bigcup_{v=t-1}M_v(t)|+|\bigcup_{v<t-\tau}M_v(t)|)
  \label{eq:SR}
\end{equation}
where $r$ is the number of global rounds and $m$ is the number of clients. $SR$ is calculated based on our lag-tolerant distribution rule (Eq. \ref{eq:lag-tol_dist}).

\begin{equation}
	VV_{SAFA} = \frac{1}{r} \sum_{t=1}^{r} 
	\textrm{var}(V_t)
  \label{eq:VV}
\end{equation}
where $V_t$ is the version distribution of trained clients at round $t$, i.e., $V_t = \{v_1, v_2,..., v_m\}$.

We change \textit{lag tolerance} (i.e., $\tau$) from 1 to 10 and set up several groups of FL tests running a regression task on the Boston Housing dataset. We set the maximum number of global rounds to 100. Apart from the best loss achieved (i.e., the minimum loss by the global model in 100 rounds), we also present the statistical results in the metrics including $EUR$, $SR$ and $VV$. 

\begin{figure}[ht]
    \centering
    \includegraphics[width=3.3in]{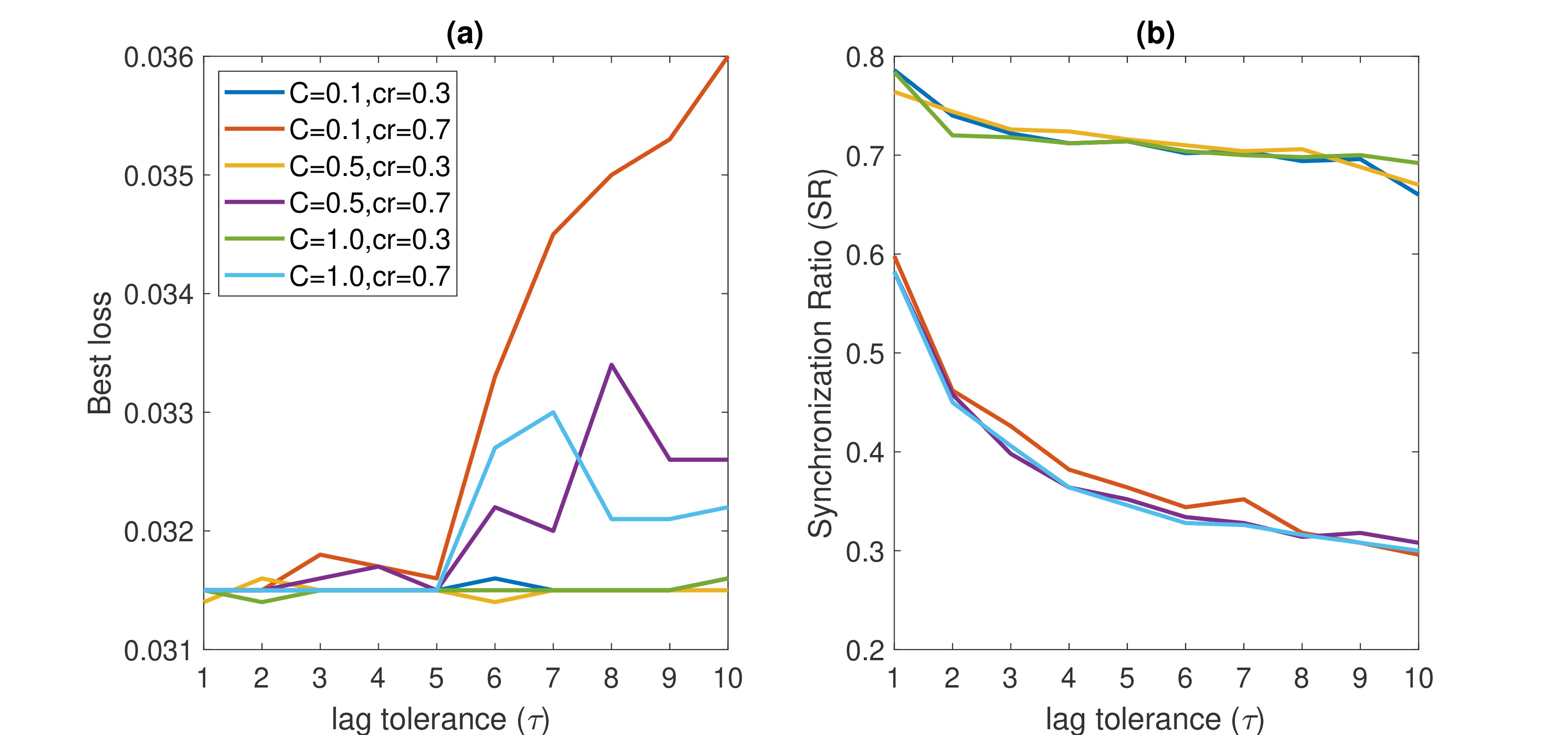}
    \caption{(a) Best loss achieved by the global model and (b) the synchronization ratio over the federated optimization with SAFA protocol under different lag tolerance settings.}
    \label{fig:loss_SR_tau}
\end{figure}

Fig. \ref{fig:loss_SR_tau}(a) draws the best loss of the global model in the FL environment where we set the selection fraction $C$ to 0.1, 0.5 and 1.0, and set the expectation of client crash probability $cr$ to 0.3 and 0.7, respectively. Fig. \ref{fig:loss_SR_tau}(b) shows the resulting synchronization ratio ($SR$). Apparently small values of \textit{lag tolerance} show a clear advantage in terms of loss. However, the overhead of communication (revealed by $SR$) is relatively large in the case where $\tau$ is set too small (e.g., 1, 2 or 3). This is expected because more clients will become deprecated and be forced to synchronize when we are less tolerant to the stragglers and stale models.

\begin{figure}[ht]
    \centering
    \includegraphics[width=3.3in]{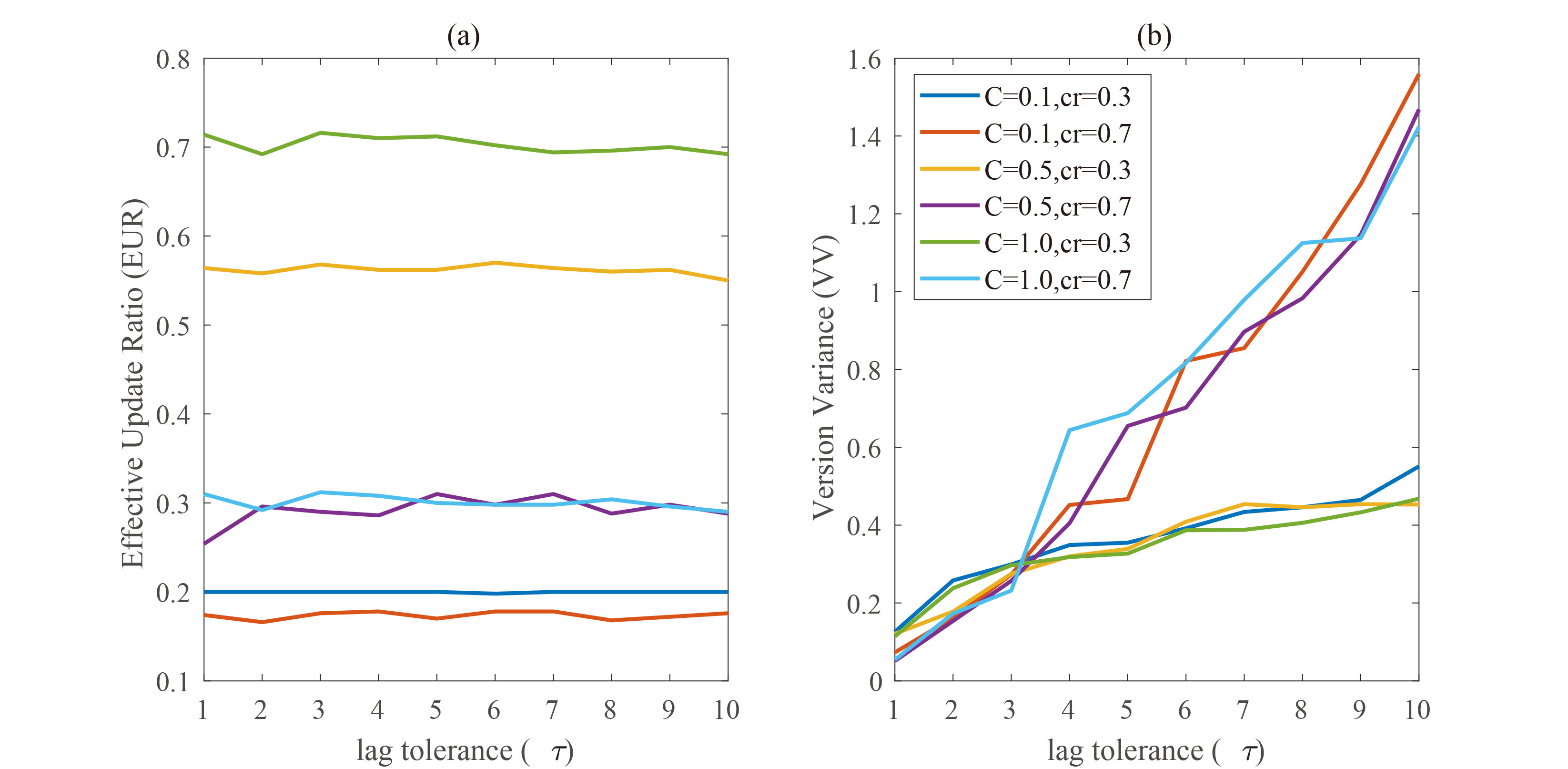}
    \caption{(a) Effective Update Ratio (EUR) and (b) Version Variance (VV) over the federated optimization with SAFA protocol under different lag tolerance settings.}
    \label{fig:EUR_VV_tau}
\end{figure}

There are multiple factors that can affect the best global model we can obtain in federated learning. We analyze it by observing the effective update ratio ($EUR$) and the variance of version ($VV$) under different FL settings -- we argue that they are two important metrics that well reflect the quality of the aggregation step, which is vital for the accuracy of the global model. From Fig. \ref{fig:EUR_VV_tau}(a) we can see $EUR$ basically remains level as the lag tolerance changes, and that $EUR$ depends on both the client fraction $C$ and the client crash probability $cr$. When $cr$ is low (e.g., $cr=0.3$), $EUR$ is slightly above the percentage quota of the clients specified by $C$, which is because of the contribution by undrafted clients. In the case of a high crash rate (e.g., $cr=0.7$), $EUR$ is restricted at a low level as it is impossible to be higher than $\mathbb{E}(|M-K|)$, which in theory is equal to $1-cr$, i.e., the portion of clients with successfully committed updates. In addition, the plot of Version Variance in Fig. \ref{fig:EUR_VV_tau}(b) reveals part of the reason why the quality of the global model degrades when \textit{lag tolerance} is set too large (see Fig. \ref{fig:loss_SR_tau}(a)). In general $VV$ increases if we make SAFA more tolerant to the stragglers (i.e., a larger value of $\tau$). It can be further observed from Fig. \ref{fig:EUR_VV_tau}(b) that as $\tau$ increases, $VV$ goes up at a much slower rate in relatively stable FL settings (e.g., $cr=0.3$) than in the extreme settings (e.g., $cr=0.7$). Combining Fig. \ref{fig:loss_SR_tau}(a) and Fig. \ref{fig:EUR_VV_tau}(b), we can see a clear correlation between $VV$ and the quality of global model especially in an unstable environment where the clients disengage frequently. 

Based on the observations, we find that a moderate \textit{lag tolerance} can largely restrain the loss of global model below a desired level and avoid the high communication cost (indicated by $SR$) in sending out the global model. Therefore we suggest setting \textit{lag tolerance} to 5 rounds in general.

\subsection{Bias Analysis}
\label{sec_bias}
In this section, we theoretically analyze the bias in client selection introduced by the discrepancy of performance and reliability between clients. Here the bias between two clients (e.g., clients A and B) refers to the ratio of client A's chance of contributing to the global model to client B's chance. It is worth mentioning that FedAvg also incurs bias (even though it uses random selection before the training starts) because the clients drop or opt out with different frequencies.

In the analysis, we consider an extreme case which represents the worst bias between the clients. In this case, clients A and B are assumed to the most and least powerful clients, respectively. Namely, clients A and B yield the shortest and longest local training time, respectively. Further, we assume they have the probabilities of dropping out in any round of training, which are denoted by $cr_A$ and $cr_B$. For the entire set of clients, we assume an overall crash ratio, denoted by $R$, which is the expected proportion of clients that drop out in a FL round. After $r$ rounds of training, the bias between clients A and B can be represented by:

\begin{equation}
	bias^{(r)} = \frac{P^{(r)}(A)}{P^{(r)}(B)}
  \label{eq:bias_def}
\end{equation}
where $P^{(r)}(A)$ (or $P^{(r)}(B)$) denotes the probability that the local update of client A (or B) is successfully aggregated in the global aggregation step in round $r$. 

We first analyze the bias generated by FedAvg (see eq. \ref{eq:bias_FedAvg}), which selects clients at the beginning of a round and the server will wait for all these selected clients to submit local updates. The local update of a selected client will always be aggregated in this round unless this client crashed. Therefore, the bias in FedAvg only depends on the clients' crash rates, which can be modeled by: 

\begin{equation}
	bias_{FedAvg}^{(r)} = \frac{1-cr_A}{1-cr_B}
  \label{eq:bias_FedAvg}
\end{equation}

In SAFA, the situation is different. \textit{C} percentage of the clients are selected from all clients that committed their local updates at the end of this round. The bias in SAFA not only depends on the crash rate, but also on the performance of the clients. A more powerful client can complete their local training faster and therefore its local update has a higher chance to be used in a round. For example, when \textit{C} percentage of the clients submit their local updates, the server will be able to finish its client selection stage and consequently the clients who fail to finish/submit before that will miss that round. 

There are two possible cases where a local update can be used in the current round: i) When a client is selected by the server, its local update will be directly applied in the current round. We denote the probability of this case by $P_D^{(r)}(A)$. ii) The local update generated by an undrafted client in last round also has the chance to be used in the current round through the bypass scheme. The probability that this case occurs is denoted by $P_S^{(r)}(A)$. Therefore, $P^{(r)}(A)$ can be calculated by summing up $P_D^{(r)}(A)$ and $P_S^{(r)}(A)$. $P^{(r)}(B)$ is decomposed similarly. 

Due to the space limitation, we only present the final expressions of $P^{(r)}(A)$ and $P^{(r)}(B)$ below in this section. The detailed derivation steps can be found in Appendix \ref{apdx:deduct}.

First, we need to consider three cases of client selection in SAFA given selection fraction $C$ and crash ratio $R$: 
\begin{itemize}
\item[] \textbf{case 1} $\iff C \geq 1-R$
\item[] \textbf{case 2} $\iff (1-C)(1-R) \leq C < 1-R$
\item[] \textbf{case 3} $\iff C < (1-C)(1-R)$. 
\end{itemize}

Literally, case 1 represents a deficit in client selection (i.e., too many crashes to fulfill the pick percentage $C$). Case 3 means that we can meet the selection ratio \textit{C} by only selecting the arrived updates from clients not selected last round since they are prioritized by SAFA. Case 2 stands between cases 1 and 3. Namely, we meet the selection ratio \textit{C} by selecting the prioritized (i.e., last-round undrafted or crashed) clients first and then other clients who also committed their local updates in this round. Considering these cases we have the following proposition:

\prop \textit{The probabilities $P^{(r)}(A)$ and $P^{(r)}(B)$ can be formulated respectively by Eqs. (\ref{eq:prA}) and (\ref{eq:prB}) given $r>1$:}

\begin{equation}
	P^{(r)}(A) = 
	\begin{cases}
		1 - cr_A 		& \text{if case 1} ,\\
		1 - cr_A		& \text{if case 2},\\
		\sigma_A^{(r-1)} - cr^2_A  & \text{otherwise}
    \end{cases}
  \label{eq:prA}
\end{equation}

\begin{equation}
	P^{(r)}(B) = 
	\begin{cases}
		1 - cr_B 					& \text{if case 1} ,\\
		\sigma_B^{(r-1)} - cr^2_B	& \text{if case 2},\\
		1 - cr_B 				 	& \text{otherwise}
    \end{cases}
  \label{eq:prB}
\end{equation}
where $\sigma^{(k)}_A  = 1-P_D^{(k)}(A)$ and $\sigma^{(k)}_B = 1-P_D^{(k)}(B)$. The proof of the proposition is detailed in Appendix \ref{apdx:deduct}. Combining the expressions of $P_D^{(r)}(A)$ and $P_D^{(r)}(B)$ (see Eqs. (\ref{eq:pDrA}) and (\ref{eq:pDrB}) in Appendix \ref{apdx:deduct}) with proper reduction, we can derive $\sigma^{(k)}_A$ and $\sigma^{(k)}_B$:

\begin{equation}
  \begin{cases}
		\sigma^{(k)}_A = \frac{2cr_A-(cr_A-1)^{k+1}-3}{cr_A-2} 	& \\
		\sigma^{(k)}_B = \frac{2cr_B-(cr_B-1)^{k+1}-3}{cr_B-2}	& \\
  \end{cases}
  \label{eq:sigma_final}
\end{equation}

Therefore, combining Eqs. (\ref{eq:prA}), (\ref{eq:prB}) and the definition of bias, we can derive the bias introduced by SAFA in round \textit{r} ($r>1$) as follows:

\begin{equation}
  bias_{SAFA}^{(r)} = 
  \begin{cases}
		\frac{1 - cr_A}{1 - cr_B}						& \text{if case 1} ,\\
		\frac{1 - cr_A}{\sigma_B^{(r-1)} - cr^2_B}	& \text{if case 2},\\
		\frac{\sigma_A^{(r-1)} - cr^2_A}{1 - cr_B} & \text{otherwise}
  \end{cases}
  \label{eq:bias_safa}
\end{equation}

%

\begin{figure}[ht]
    \centering
    \includegraphics[width=2.3in]{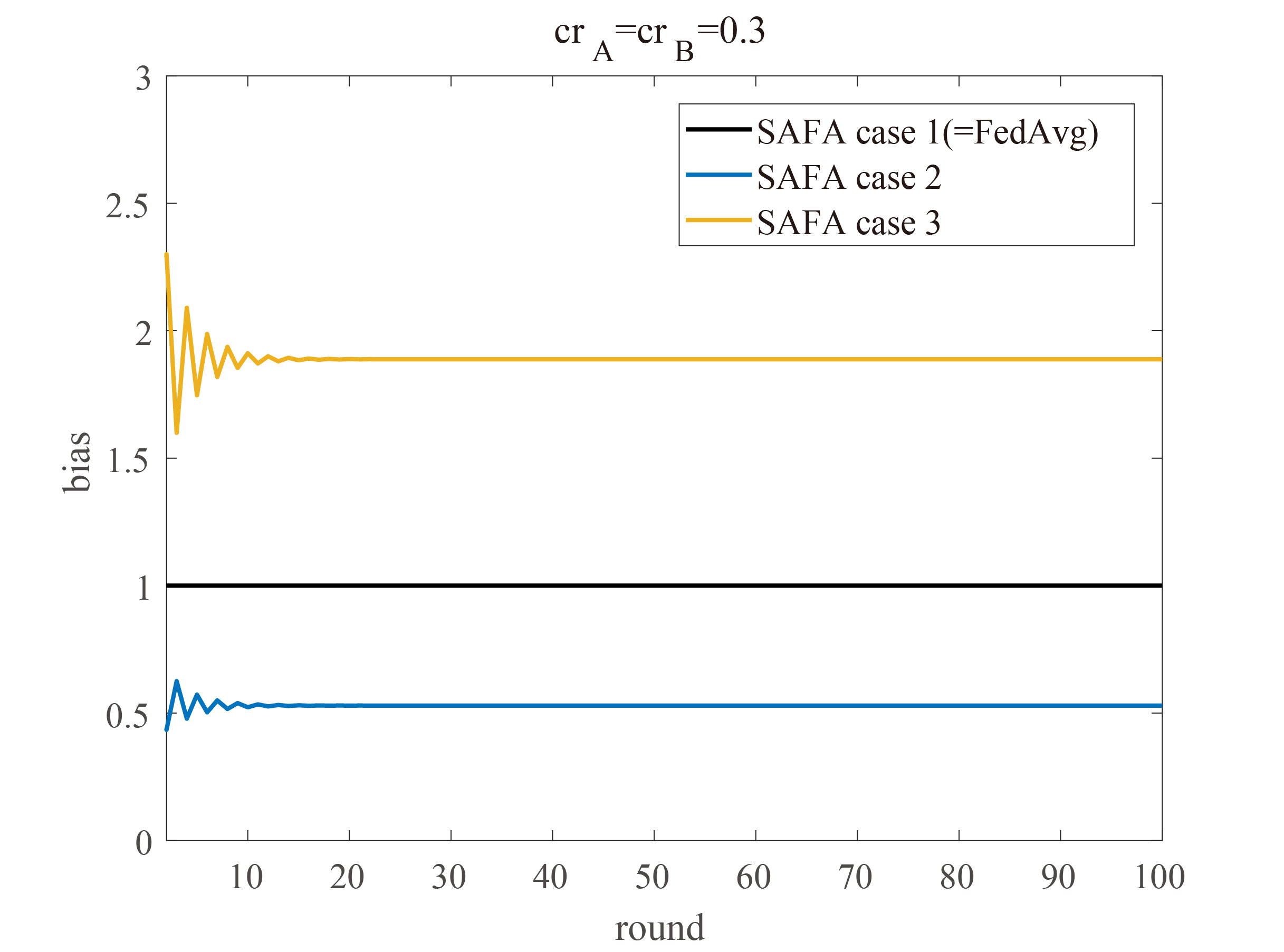}
    \caption{The bias incurred by FedAvg and SAFA (in the circumstances of three different cases) as a function of federated round index. Here both clients A and B have the same crash rate of 0.3, and the results are similar when setting different $cr_A$ and $cr_B$ according to our experiments.}
    \label{fig:bias_plot}
\end{figure}

Fig. \ref{fig:bias_plot} visualizes the bias of FedAvg and SAFA as a function of round index $r$. In case 1 where all local updates committed by the clients are aggregated, we have a fixed bias of $\frac{1 - cr_A}{1 - cr_B}$, which is the same as FedAvg. In case 2, client B, as the slowest one, will be picked (once it has committed) by the server as long as it was undrafted or crashed in the previous round, which effectively reduces the bias to a level below that of FedAvg. As for case 3, the quota (decided by $C$) will be fulfilled only with last-round undrafted or crashed clients. Assuming both A and B missed last round, client B is disadvantageous because the server is likely to end the round before B finishes training when the fraction $C$ has been fulfilled by other faster clients (including client A). In all these cases, the bias between A and B converges after a few rounds once FL starts.

\section{Experimental Evaluation}

\subsection{Experiment Setup}
We conducted extensive experiments to evaluate the effectiveness of the SAFA protocol on three typical machine learning tasks. \textit{Task 1} is to fit a regression model on the public Boston Housing dataset\footnote{\url{https://www.cs.toronto.edu/~delve/data/boston/bostonDetail.html}}, which is available in public repositories. \textit{Task 2} is to learn a handwritten digit image classification model implemented using a convolutional neural network (CNN), which is comprised of two 5x5 convolution layers (the first one with 20 channels and the second with 50 channels) with 2x2 max pooling, a fully-connected layer with ReLu as the activation function, and a final softmax output layer. This light-weight CNN is suitable for end devices with small memory size and also adopted in the experiment by \cite{origin_FL}. \textit{Task 3} is to learn a classification model for detecting network intrusion given the TCP dump data. For this task we extract the TCP-connection examples from the KDD Cup'99 dataset\footnote{\url{https://kdd.ics.uci.edu/databases/kddcup99/kddcup99.html}} and use Support Vector Machine (SVM) as the classification model. 

We set up separate environments for these three learning tasks to investigate the performance of our protocol in different FL settings. To simulate the unreliability of clients, we set a crash probability ($cr$) in each run of test and assume each client has the equal chance $cr$ to drop out in any round of federated training. For a given task, we use identical local training settings (e.g., mini-batch size) for all the clients and use identical global settings (e.g., the maximum number of rounds and round time limit) for each protocol. The details of the experiment setup is shown in Table \ref{tab:exp_setup}.

\begin{table}[ht]
\centering
\caption{Experimental setup for federated learning}
\begin{tabular}{ l l l l l } 
 \hline
 parameter 	  	 	& symbol	&Task 1 		&Task 2 &Task 3 	\\ 
 \hline
 dataset		 	& $D$		&Boston			&MNIST	&KDDCup99	\\
 \# of features	 	& $d$		&13				&28x28	&35			\\
 model				& $w$		&Regression		&CNN	&SVM		\\
 dataset size	 	& $n$		&506			&70k	&186k		\\
 \# of clients	 	& $m$		&5				&100	&500		\\
 max \# of rounds 	& $R$		&100			&50		&100		\\
 \# of local epochs & $E$		&3				&5		&5			\\
 mini-batch size 	& $B$		&5				&40		&100		\\
 learning rate 		& $lr$		&1e-4			&1e-3	&1e-2		\\
 \hline
\end{tabular}
\label{tab:exp_setup}
\end{table}

To simulate data imbalance and the heterogeneity in end devices, we assume the size of data partitions (i.e., local data size) follows the Gaussian distribution $\mathcal{N}(\mu, 0.3\mu)$ where $\mu=n/m$, and assume clients' performance follows the exponential distribution with $\lambda=1.0$. Here we define the performance of a client as the number of batches it can process per second in training. End devices (i.e., clients) may be unreliable and crash occasionally by a probability of $\rho_k$. In the experiment we assume clients crash independently with the same probability in any federated round and set $\rho_k$ to be $cr$, i.e. $\rho_k=cr, k=1,2,...,m$.

For comparison, we also implemented FedAvg \cite{origin_FL}, FedCS and a fully local training process as the baselines. FedCS \cite{FedCS} is a refined FL protocol that has to estimate the speed that clients work and filters out some slow clients proactively (at the stage of client selection) to improve the overall efficiency of FL. The fully local protocol never performs the global aggregation until the end of the final round.

\subsection{Results}
\begin{figure}[ht]
    \centering
    \includegraphics[width=3.3in]{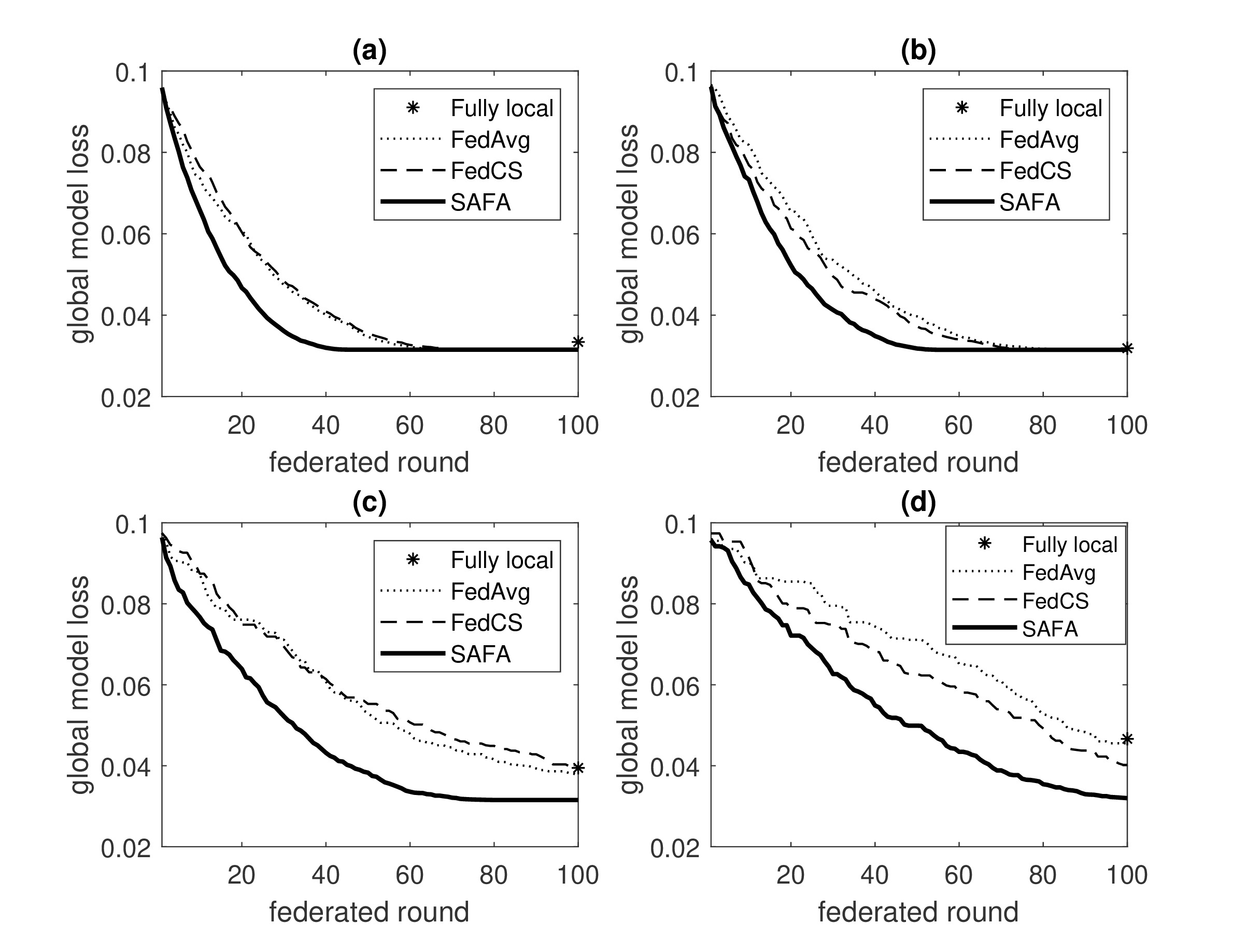}
    \caption{The loss trace of the global model as the FL process progresses on Task 1 where the client fraction is set to 0.3 and the crash probability is set to 0.1, 0.3 ,0.5 and 0.7 for the four sub-figures (a)-(d), respectively}
    \label{fig:task1_loss}
\end{figure}

\begin{figure}[ht]
    \centering
    \includegraphics[width=3.3in]{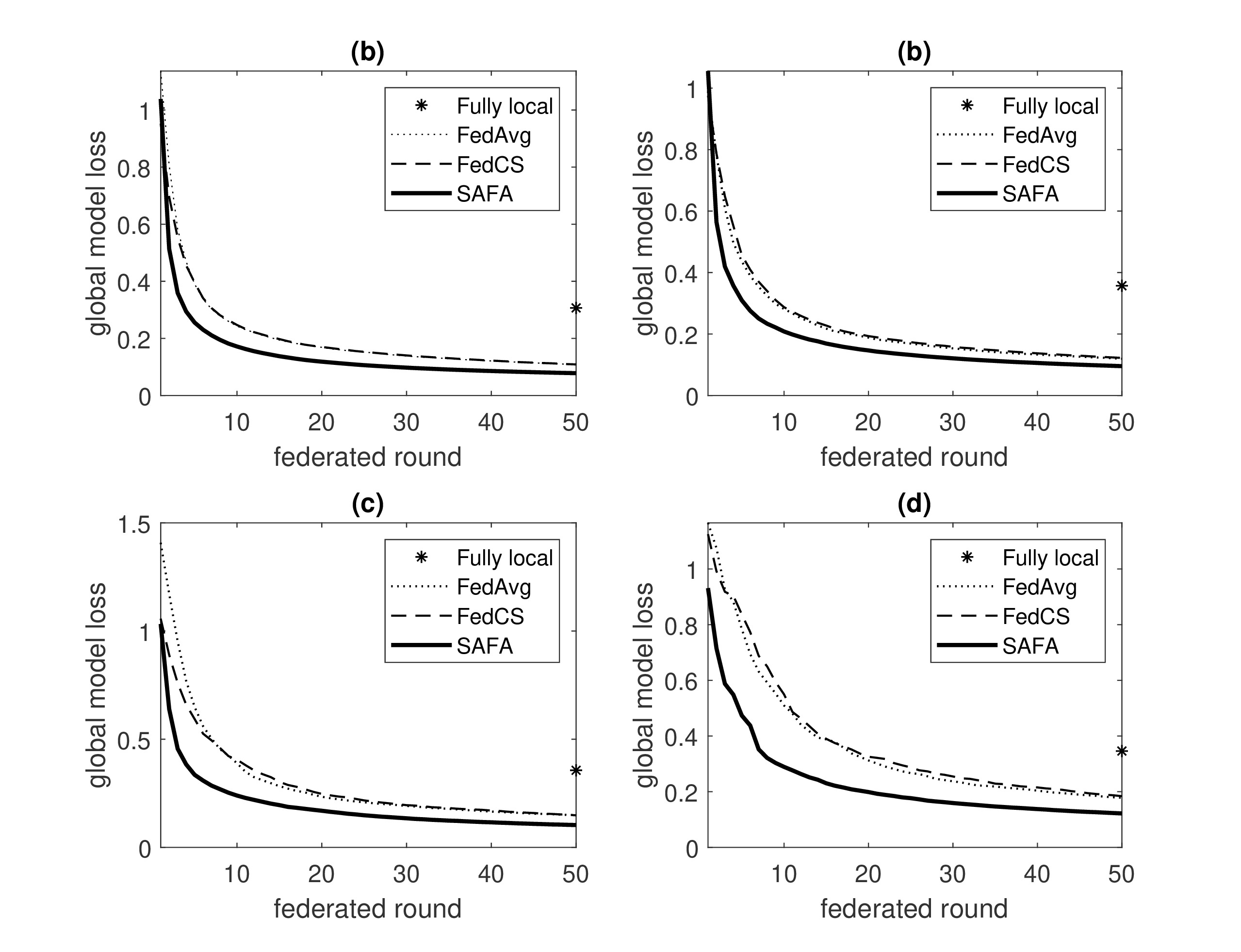}
    \caption{The loss trace of the global model as the FL process progresses on Task 2 where the client fraction is set to 0.3 and the crash probability is set to 0.1, 0.3 ,0.5 and 0.7 for the four sub-figures (a)-(d), respectively.}
    \label{fig:task2_loss}
\end{figure}

\begin{figure}[ht]
    \centering
    \includegraphics[width=3.3in]{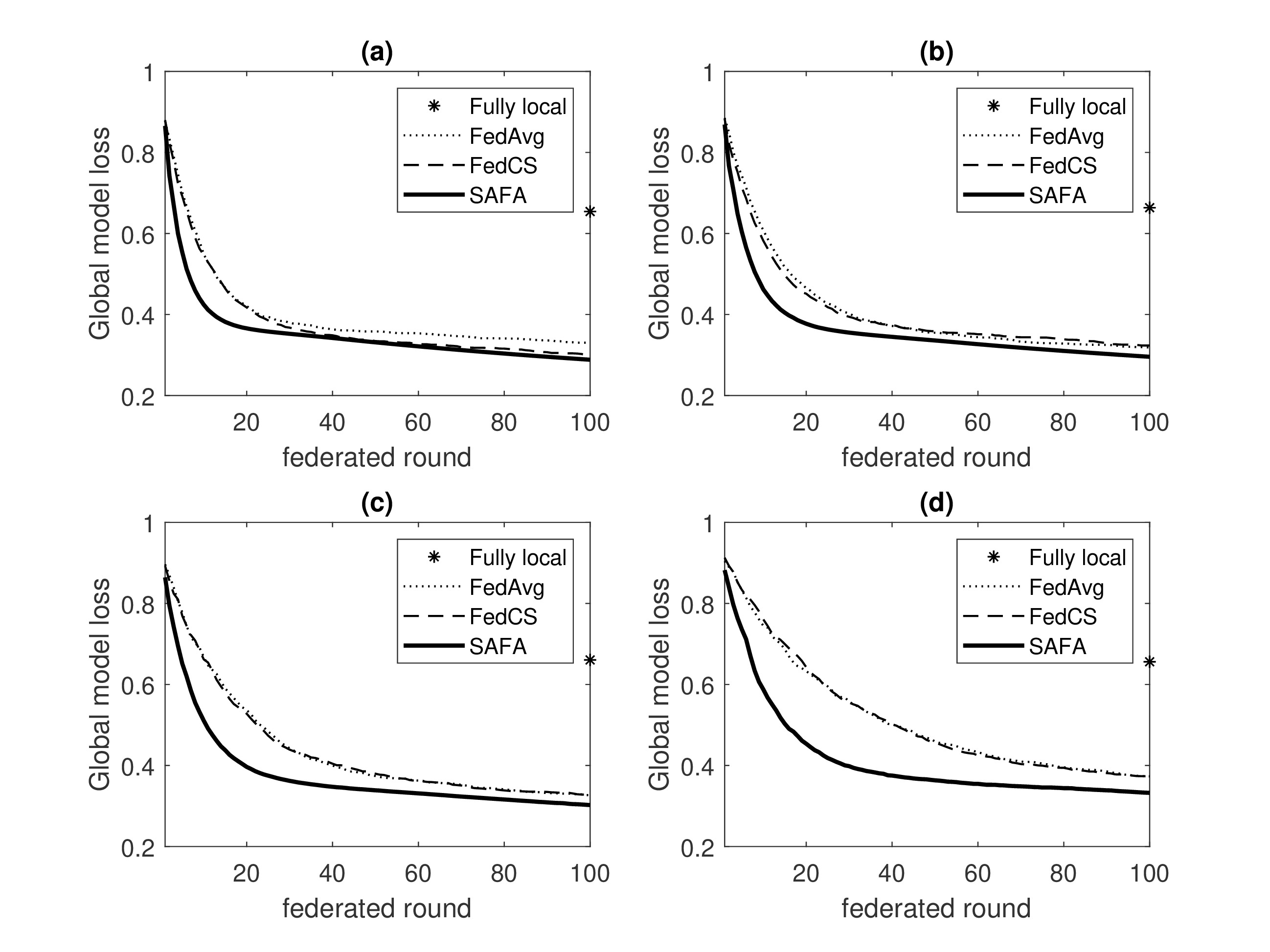}
    \caption{The loss trace of the global model as the FL process progresses on Task 3, where the client fraction is set to 0.3 and the crash probability is set to 0.1, 0.3 ,0.5 and 0.7 for the four sub-figures (a)-(d) respectively}
    \label{fig:task3_loss}
\end{figure}

In this section we present the results of our experiments and discuss the evaluated FL protocols in terms of the quality of the obtained global model (shown in Figs. \ref{fig:task1_loss}, \ref{fig:task2_loss} and \ref{fig:task3_loss}, with more details in Tables \ref{tab:task1_acc}, \ref{tab:task2_acc} and \ref{tab:task3_acc}) as well as holistic metrics including round efficiency (summarized in Tables \ref{tab:task1_length}, \ref{tab:task2_length} and \ref{tab:task3_length}), communication overheads (in Tables \ref{tab:task1_Tdist}, \ref{tab:task2_Tdist} and \ref{tab:task3_Tdist}) and local resource utilization (Tables \ref{tab:task1_SR_futile}, \ref{tab:task2_SR_futile} and \ref{tab:task3_SR_futile}). 

A main objective of our work is to boost the round efficiency (i.e., reducing the average length of a federated round), the convergence rate and the resulting accuracy (of the global model). In our experiments, we measure the length of a federated round by considering both local training time and communication overheads, which is captured by Eq. (\ref{eq:round_length}).

\begin{equation}
	T = \min \big\{T_{lim},\, T_{dist} + \max_k \{T^{down}_k + T^{up}_k +T^{train}_k\} \big\}
  \label{eq:round_length}
\end{equation}
where $T_{lim}$ is the preset upper limit of round length. $T^{train}_k$, $T^{down}_k$ and $T^{up}_k$ denote local training time, model download and upload time for client $k$, respectively. $T^{down}_k$ and $T^{up}_k$ depend on model size and device bandwidth. Using a local network setting similar to that in \cite{FedCS}, we assign a stable bandwidth of 1.40Mbps to each client. For client $k$, its local training time (i.e., $T^{train}_k$) is determined using Eq. (\ref{eq:T_train}):

\begin{equation}
	T^{train}_k = \frac{|B_k| \cdot E}{s_k}
  \label{eq:T_train}
\end{equation}
where $E$ is the number of local epochs and $|B_k|$ is the number of batches on device $k$ ($|B_k|$ depends on the size of its local data partition and the preset batch size). In the experiment, a client's performance is defined as the number of batches the client is able to process per second. $s_k$ denotes the performance of client $k$. We assume that clients' performance follows the exponential distribution with $\lambda=1.0$.

For a federated round, $T_{dist}$ denotes the server-side overhead for distributing the global model to the end devices. In this paper we assume the server can fully utilize its bandwidth to send models in parallel via intermediate network elements \cite{FL_WangS} to the clients. Thus $T_{dist}$ depends on the number of model copies to distribute (denoted by $m_{sync}$) and the communication bandwidth of the server (denoted by $bw$). $T_{dist}$ is formulated in Eq. (\ref{eq:T_dist}). Given a FL protocol, $T_{dist}$ of a round is closely correlated to its Synchronization Ratio. The increase in SR indicates a higher average communication cost at the stage of model distribution.

\begin{equation}
	T_{dist} = \frac{m_{sync} \cdot model\_size}{bw}
  \label{eq:T_dist}
\end{equation}
where the server bandwidth $bw$ is set to 10Gbps in our experiment considering the prevailing 10-Gigabit Ethernet connection. Models are usually compressed before transmission. We use 10MB as the model size following the result presented in \cite{10MB_compress}. 

For different machine learning models, we define their accuracy in different ways, as shown in Table \ref{tab:acc_def}. In the table, $y$ and $\hat{y}$ denote the label and the output of the model, respectively. The function $\phi(\cdot)$ returns 1 if $\hat{y}$ matches $y$, otherwise it returns 0.

\begin{table}[ht]
\centering
\caption{Formulating accuracy of the global model for the three tasks}
\begin{tabular}{ l l } 
 \hline
 ML task 	  	 	& accuracy formulation\\ 
 \hline
 Task 1: regression	& $acc = 1-\frac{1}{n}\sum_{i=1}^n{\frac{|y_i-\hat{y_i}|}{\mathrm{max}(y_i,\hat{y_i})}}$ \\
 Task 2: CNN		& $acc = \frac{1}{n}\sum_{i=1}^n\phi(y_i,\hat{y_i})$ \\
 Task 3: SVM		& $acc = \frac{1}{n}\sum_{i=1}^n\mathrm{max}(0, \mathrm{sign}(y_i\cdot \hat{y_i}))$ \\
 \hline
\end{tabular}
\label{tab:acc_def}
\end{table}

\noindent \emph{Task 1: Regression} 

In this task, we aim to learn a regression model on a small group of clients to predict the median value of a house in the area of Boston Mass. Input features include 13 properties about the estate such as average number of rooms per dwelling and crime rate. In this experiment, we ran FL with every candidate protocol (i.e., SAFA, FedAvg, FedCS and fully local training) and compare their effectiveness in terms of the achieved accuracy of the global model, round efficiency and communication overhead.

\begin{table}[ht]
\centering
\caption{Average length of a federated round in secs on Task 1 wherein each protocol was tested with varying selection fraction under different environment settings. Round time limit is set to 830s considering the client performance and data distribution.}
\begin{tabular}{l l l l l l} 
 \hline
 \multicolumn{6}{c}{Avg. round length (Task 1: regression)}\\
 \multicolumn{6}{c}{FedAvg}\\
 $cr$	& $C=0.1$	&$C=0.3$	&$C=0.5$ 	&$C=0.7$	&$C=1.0$	\\
 \hline
 0.1	&316.22		&489.37		&586.90		&731.12		&808.59		\\
 0.3	&429.63		&652.39		&641.40		&736.53		&832.02		\\
 0.5	&372.43		&495.37		&475.14		&621.91		&676.41		\\
 0.7	&354.34		&405.86		&593.10		&728.25		&661.67		\\
 \hline
 \multicolumn{6}{c}{FedCS}\\
 $cr$	&$C=0.1$	&$C=0.3$	&$C=0.5$ 	&$C=0.7$	&$C=1.0$	\\
 \hline
 0.1	&207.50		&487.47		&564.20		&656.49		&786.96		\\
 0.3	&336.97		&519.58		&651.23		&401.95		&832.02		\\
 0.5	&186.51		&221.46		&467.98		&621.91		&676.41		\\
 0.7	&195.09		&398.81		&584.68		&393.09		&661.67		\\
 \hline
 \multicolumn{6}{c}{SAFA}\\
 $cr$	&$C=0.1$	&$C=0.3$	&$C=0.5$ 	&$C=0.7$	&$C=1.0$	\\
 \hline
 0.1	&149.69		&389.44		&540.41		&606.48		&734.40		\\
 0.3	&202.44		&430.68		&583.22		&371.77		&699.23		\\
 0.5	&169.33		&215.66		&408.85		&510.85		&508.23		\\
 0.7	&161.81		&293.09		&402.18		&411.06		&379.29		\\
 \hline						
\end{tabular}
\label{tab:task1_length}
\end{table}

It can be seen from Fig. \ref{fig:task1_loss} and Table \ref{tab:task1_acc} (in Appendix \ref{apdx:tabs}) that SAFA significantly improves the convergence rate as well as the best accuracy achieved by the global regression model, especially under settings of unstable environments (i.e., $cr \geq 0.5$). This is mainly attributed to our staleness-tolerant mechanism. Another advantage of the tolerance to stragglers is the preservation of local training results. We use the metric \textit{Futility Percentage} to measure the percentage of local progress that is wasted due to the model synchronization forced by the server (FedAvg and FedCS force the selected clients to overwrite its local model with the latest global model). Results of SR and futility percentage in Table \ref{tab:task1_SR_futile} show that the wasted training progress is reduced by SAFA effectively.

\begin{table}[ht]
\centering
\caption{Average model distribution overhead (unit: seconds) on Task 1}
\begin{tabular}{l l l l l l} 
 \hline
 \multicolumn{6}{c}{Avg. $T_{dist}$ (Task 1: regression)}\\
 \multicolumn{6}{c}{FedAvg}\\
 $cr$	& $C=0.1$	&$C=0.3$	&$C=0.5$ 	&$C=0.7$	&$C=1.0$	\\
 \hline
 0.1	&0.40		&0.81		&1.21		&1.62		&2.02		\\
 0.3	&0.40		&0.81		&1.21		&1.62		&2.02		\\
 0.5	&0.40		&0.81		&1.21		&1.62		&2.02		\\
 0.7	&0.40		&0.81		&1.21		&1.62		&2.02	\\
 \hline
 \multicolumn{6}{c}{FedCS}\\
 $cr$	&$C=0.1$	&$C=0.3$	&$C=0.5$ 	&$C=0.7$	&$C=1.0$	\\
 \hline
 0.1	&0.33		&0.81		&1.21		&1.62		&2.02		\\
 0.3	&0.40		&0.81		&1.21		&1.31		&2.02		\\
 0.5	&0.33		&0.64		&1.21		&1.62		&2.02		\\
 0.7	&0.33		&0.81		&1.21		&1.29		&2.02		\\
 \hline
 \multicolumn{6}{c}{SAFA}\\
 $cr$	&$C=0.1$	&$C=0.3$	&$C=0.5$ 	&$C=0.7$	&$C=1.0$	\\
 \hline
 0.1	&1.84		&1.83		&1.80		&1.84		&1.81		\\
 0.3	&1.49		&1.46		&1.43		&1.40		&1.41		\\
 0.5	&1.00		&1.07		&0.96		&1.05		&1.02		\\
 0.7	&0.76		&0.69		&0.77		&0.75		&0.74		\\
 \hline						
\end{tabular}
\label{tab:task1_Tdist}
\end{table}

As shown in Tables \ref{tab:task1_length} and \ref{tab:task1_Tdist}, there is not much difference in average round length and model distribution overhead due to the very limited number of devices used to run task 1. But we still observed notable efficiency boost and convergence speedup by SAFA under the circumstance where the selection fraction $C$ is very small. With $C$ set to 0.1, SAFA halves the time required to finish a federated round compared to FedAvg.

\bigskip
\noindent \emph{Task 2: CNN}

We divided the MNIST dataset into $m$ partitions of which the sizes are random variables (following Gaussian distribution). The CNN models with randomly initialized weights are created on 100 clients and we again tested Fully local, FedAvg, FedCS and SAFA under a variety of FL settings.

\begin{table}[ht]
\centering
\caption{Average length of a federated round in secs on Task 2 wherein each protocol was tested with varying selection fraction under different environment settings. Round time limit is set to 5600s considering the client performance and data distribution.}
\begin{tabular}{l l l l l l} 
 \hline
 \multicolumn{6}{c}{Avg. round length (Task 2: CNN)}\\
 \multicolumn{6}{c}{FedAvg}\\
 $cr$	& $C=0.1$	&$C=0.3$	&$C=0.5$ 	&$C=0.7$	&$C=1.0$	\\
 \hline
 0.1	&3402.55	&5557.25	&5610.20	&5614.28	&5620.40	\\
 0.3	&5410.97	&5606.12	&5610.20	&5614.28	&5620.40	\\
 0.5	&5602.04	&5606.12	&5610.20	&5614.28	&5620.40	\\
 0.7	&5602.04	&5606.12	&5610.20	&5614.28	&5620.40	\\
 \hline
 \multicolumn{6}{c}{FedCS}\\
 $cr$	&$C=0.1$	&$C=0.3$	&$C=0.5$ 	&$C=0.7$	&$C=1.0$	\\
 \hline
 0.1	&1487.96	&2133.02	&3668.70	&1871.65	&1982.91	\\
 0.3	&1261.59	&1542.61	&3132.86	&2349.46	&5395.54	\\
 0.5	&1273.37	&1642.59	&3025.75	&2876.63	&3162.02	\\
 0.7	&1253.74	&1969.28	&2180.46	&4344.88	&2530.01	\\
 \hline
 \multicolumn{6}{c}{SAFA}\\
 $cr$	&$C=0.1$	&$C=0.3$	&$C=0.5$ 	&$C=0.7$	&$C=1.0$	\\
 \hline
 0.1	&198.28		&315.33		&3703.81	&1708.93	&1947.90	\\
 0.3	&206.88		&368.01		&2691.25	&1899.23	&2149.23	\\
 0.5	&203.48		&800.64		&2573.60	&2727.25	&2186.67	\\
 0.7	&241.86		&1893.14	&1877.30	&2619.79	&2340.80	\\
 \hline						
\end{tabular}
\label{tab:task2_length}
\end{table}

\begin{table}[ht]
\centering
\caption{Average model distribution overhead (unit: seconds) on Task 2}
\begin{tabular}{l l l l l l} 
 \hline
 \multicolumn{6}{c}{Avg. $T_{dist}$ (Task 2: CNN)}\\
 \multicolumn{6}{c}{FedAvg}\\
 $cr$	& $C=0.1$	&$C=0.3$	&$C=0.5$ 	&$C=0.7$	&$C=1.0$	\\
 \hline
 0.1	&2.04		&6.12		&10.20		&14.28		&20.40		\\
 0.3	&2.04		&6.12		&10.20		&14.28		&20.40		\\
 0.5	&2.04		&6.12		&10.20		&14.28		&20.40		\\
 0.7	&2.04		&6.12		&10.20		&14.28		&20.40		\\
 \hline
 \multicolumn{6}{c}{FedCS}\\
 $cr$	&$C=0.1$	&$C=0.3$	&$C=0.5$ 	&$C=0.7$	&$C=1.0$	\\
 \hline
 0.1	&2.04		&6.12		&10.20		&14.14		&20.40		\\
 0.3	&2.02		&6.05		&10.20		&14.13		&20.40		\\
 0.5	&2.04		&6.06		&10.20		&14.28		&20.20		\\
 0.7	&2.04		&6.12		&10.11		&14.28		&20.20		\\
 \hline
 \multicolumn{6}{c}{SAFA}\\
 $cr$	&$C=0.1$	&$C=0.3$	&$C=0.5$ 	&$C=0.7$	&$C=1.0$	\\
 \hline
 0.1	&18.27		&18.45		&18.26		&18.47		&18.38		\\
 0.3	&14.45		&14.65  	&14.48		&14.54		&14.69		\\
 0.5	&10.89		&10.51		&10.70		&10.84		&10.58		\\
 0.7	&7.17		&7.23		&7.55		&7.21		&7.41		\\
 \hline						
\end{tabular}
\label{tab:task2_Tdist}
\end{table}

As a result, the Fully Local protocol can finish with an accuracy around 90\% on this classification task with the CNN model, while FedAvg, FedCS and SAFA raise that to $96.0\% \sim 98.0\%$ (Table \ref{tab:task2_acc} in Appendix \ref{apdx:tabs}). SAFA shows a significant advantage in round efficiency (see Table \ref{tab:task2_length}) - it is able to achieve up to $27\times$ and $6\times$ speed-up compared to FedAvg and FedCS in an unreliable environment where clients frequently opt/drop out and only a small fraction (i.e., $C=$ 0.1 or 0.3) of them are allowed to participate in a round.

The average $T_{dist}$ for SAFA mainly depends on client crash probability (see Table \ref{tab:task2_Tdist}), and it remains at a low level with $cr \geq 0.5$. In the case where devices are more reliable in local training (i.e., $cr < 0.5$), SAFA embraces a greater number of updates and results in a slightly higher cost (of tens of seconds) during the stage of model distribution, but the overhead is still acceptable considering the overall length of a federated round (which could last thousands of seconds, see Table \ref{tab:task2_length}).

\bigskip
\noindent \emph{Task 3: SVM}

For this task we use a relatively large data set containing 186,480 TCP dump records including several types of network intrusions. The target is to learn a global SVM model to recognize malicious connections and normal connections. We dispersed the dataset onto 500 clients to perform FL with SAFA and other existing training protocols.

\begin{table}[ht]
\centering
\caption{Average length of a federated round in secs on Task 3 wherein each protocol was tested with varying selection fraction under different environment settings. Round time limit is set to 1620s considering the client performance and data distribution.}
\begin{tabular}{l l l l l l} 
 \hline
 \multicolumn{6}{c}{Avg. round length (Task 3: SVM)}\\
 \multicolumn{6}{c}{FedAvg}\\
 $cr$	& $C=0.1$	&$C=0.3$	&$C=0.5$ 	&$C=0.7$	&$C=1.0$	\\
 \hline
 0.1	&1640.20	&1680.60	&1721.00	&1761.40	&1822.00	\\
 0.3	&1640.20	&1680.60	&1721.00	&1761.40	&1822.00	\\
 0.5	&1640.20	&1680.60	&1721.00	&1761.40	&1822.00	\\
 0.7	&1640.20	&1680.60	&1721.00	&1761.40	&1822.00	\\
 \hline
 \multicolumn{6}{c}{FedCS}\\
 $cr$	&$C=0.1$	&$C=0.3$	&$C=0.5$ 	&$C=0.7$	&$C=1.0$	\\
 \hline
 0.1	&788.75		&1319.17	&1607.42	&1539.14	&1802.09	\\
 0.3	&685.26		&1216.12	&1521.82	&1617.97	&1775.50	\\
 0.5	&714.73		&1229.87	&1371.03	&1605.23	&1821.60	\\
 0.7	&754.52		&1190.44	&1526.23	&1573.42	&1731.65	\\
 \hline
 \multicolumn{6}{c}{SAFA}\\
 $cr$	&$C=0.1$	&$C=0.3$	&$C=0.5$ 	&$C=0.7$	&$C=1.0$	\\
 \hline
 0.1	&310.70		&353.98		&1419.29	&1514.38	&1802.15	\\
 0.3	&274.03		&330.32		&1499.79	&1559.50	&1762.51	\\
 0.5	&242.93		&398.27		&1317.91	&1476.14	&1724.52	\\
 0.7	&212.52		&1187.96	&1313.99	&1223.72	&1690.61	\\
 \hline						
\end{tabular}
\label{tab:task3_length}
\end{table}

Table \ref{tab:task3_acc} (in Appendix \ref{apdx:tabs}) shows that FedAvg, FedCS and SAFA can produce very accurate global models (with the classification accuracy of over $99\%$) after convergence. SAFA could incur higher overhead in model distribution (as the SR is larger for SAFA, see Tables \ref{tab:task3_Tdist} and \ref{tab:task3_SR_futile} in some cases). Nevertheless, SAFA still significantly outperforms FedAvg and FedCS by $7.7\times$ and $3.7\times$, respectively, in average round length (see Table \ref{tab:task3_length}). Its advantage decreases as more clients are set to engage in training but it is still the most efficient protocol. In contrast to FedAvg and FedCS, SAFA capitalizes the contribution from straggling clients effectively, leading to a very small futility percentage (below 4\%, see Table \ref{tab:task3_SR_futile} in Appendix \ref{apdx:tabs}) on this task, which means that the majority of local training progresses make contribution to the convergence of the final global model, even in a very unreliable environment. 

\begin{table}[ht]
\centering
\caption{Average model distribution overhead (in seconds) on Task 3}
\begin{tabular}{l l l l l l} 
 \hline
 \multicolumn{6}{c}{Avg. $T_{dist}$ (Task 3: SVM)}\\
 \multicolumn{6}{c}{FedAvg}\\
 $cr$	& $C=0.1$	&$C=0.3$	&$C=0.5$ 	&$C=0.7$	&$C=1.0$	\\
 \hline
 0.1	&20.20		&60.60		&101.00		&141.40		&202.00		\\
 0.3	&20.20		&60.60		&101.00		&141.40		&202.00		\\
 0.5	&20.20		&60.60		&101.00		&141.40		&202.00		\\
 0.7	&20.20		&60.60		&101.00		&141.40		&202.00		\\
 \hline
 \multicolumn{6}{c}{FedCS}\\
 $cr$	&$C=0.1$	&$C=0.3$	&$C=0.5$ 	&$C=0.7$	&$C=1.0$	\\
 \hline
 0.1	&20.20		&60.48		&100.78		&140.79		&201.60		\\
 0.3	&20.11		&60.60		&100.81		&141.09		&201.60		\\
 0.5	&20.13		&60.60		&100.85		&140.84		&201.60		\\
 0.7	&20.20		&60.60		&100.61		&141.14		&201.19		\\
 \hline
 \multicolumn{6}{c}{SAFA}\\
 $cr$	&$C=0.1$	&$C=0.3$	&$C=0.5$ 	&$C=0.7$	&$C=1.0$	\\
 \hline
 0.1	&181.95		&182.32		&181.49		&181.84		&182.15		\\
 0.3	&142.89		&141.91		&141.95		&142.50		&142.81		\\
 0.5	&104.38		&104.56		&105.34		&104.59		&104.52		\\
 0.7	&70.62		&70.63		&70.55		&70.05		&70.61		\\
 \hline						
\end{tabular}
\label{tab:task3_Tdist}
\end{table}

\subsection{Discussion}
The experimental results with several tasks including regression and classification demonstrate the effectiveness of applying our semi-asynchronous protocol to FL with unreliable clients. The improvement achieved by SAFA lies in three-fold: i) faster convergence of the global model and a higher accuracy achieved, ii) significant reduction in average round length, and iii) increased utilization of local training progress made by the stragglers. A few interesting phenomena were also observed in our experiments. First and foremost, we find that increasing the client fraction $C$ does not always improve the quality of the global model. For example, a reasonably high accuracy is obtained by setting $C$ to 0.3 or 0.5 (instead of 1.0) in task 2 in the case of a low crash probability. This in some ways infers that involving more clients each round is not always beneficial (or has very limited benefit). In addition, we notice that fully local training without round-wise aggregation is in some cases able to produce a reasonably good model, e.g., in the cases of Task 1 with $C=0.3$ and Task 3 with $C=0.1$ and $cr=0.7$. Also, we find that the synchronous FL protocol FedAvg can produce a global model slightly better than our solution in the case of $C=1.0$, i.e., trying to involve all clients in every round. This advantage is probably brought by the feature that pure synchronization can avoid the negative effect from stale models, which amplifies as a larger fraction of clients get involved. However, it is practically unrealistic to set a big $C$ for FL because communication could be expensive while the enhancement of the resulting accuracy is very limited (see Tables \ref{tab:task1_acc}, \ref{tab:task2_acc} and \ref{tab:task3_acc} in Appendix \ref{apdx:tabs}).

\section{Conclusion}
Aiming at improving the efficiency of federated learning with unreliable end devices, we propose a semi-asynchronous protocol which incorporates a novel client selection algorithm decoupling the central server and the selected clients for a reduction of average round time as well as a lag-tolerant mechanism in model distribution for tackling the tradeoff between faster convergence and lower communication overhead. We also analyze the upper bound of the bias introduced by using SAFA in FL. The results of experimental evaluation on three typical machine learning tasks show that our protocol effectively enhances the round efficiency of federated optimization process, improves the quality of the global model and reduces local resource wastage at a relatively low cost of communication.

Considering the subtle correlation between local models and the global model, we plan to look into the balance between generating the best local models for end devices and obtaining an optimal global model in the central server. As another part of future work, we are also going to investigate how to further improve federated learning using model parallelism and compression.

\section*{Acknowledgement}
This work is partially supported by Worldwide Byte Security Information Technology Co. LTD, National Natural Science Foundation of China (Grant No. 61772205), Guangzhou Development Zone Science and Technology(Grant No. 2018GH17), Major Program and of Guangdong Basic and Applied Research (2019B030302002), Guangdong project (Grant Nos. 2017B030314073, 2018B030325002), the EPSRC Centre for Doctoral Training in Urban Science (under EPSRC grant no. EP/L016400/1), the Alan Turing Institute (under EPSRC grant EP/N510129/1 and PETRAS), and the National Center of Excellence for IoT Systems Cybersecurity (under grant EP/S035362/1).

\newpage

%

\appendices
\section{Deduction of $P^{(r)}(A)$ and $P^{(r)}(B)$}
\label{apdx:deduct}
\prop \textit{Both $P^{(r)}(A)$ and $P^{(r)}(B)$ can be decomposed as the summation of two probabilities:}

\begin{equation}
	P^{(r)}(A) = P_D^{(r)}(A) + P_S^{(r)}(A)
  \label{eq:decomp_PA}
\end{equation}

\begin{equation}
	P^{(r)}(B) = P_D^{(r)}(B) + P_S^{(r)}(B)
  \label{eq:decomp_PB}
\end{equation}
where $P_D^{(r)}(A)$ denotes the probability by which the update from client A goes directly into the cache, and $P_S^{(r)}(A)$ that client A's model in the bypass structure goes into the cache in round $r$. $P_D^{(r)}(B)$ and $P_S^{(r)}(B)$ are defined in a similar way. For the three cases considered, for client A we have:

\begin{equation}
	P_D^{(r)}(A) = 
	\begin{cases}
		1 - cr_A 					& \text{if case 1} ,\\
		1 - cr_A					& \text{if case 2},\\
		(1-cr_A)(1-P_D^{(r-1)}(A)) 	& \text{otherwise}
    \end{cases}
  \label{eq:pDrA}
\end{equation}

\begin{equation}
	P_S^{(r)}(A) = 
	\begin{cases}
		0 		& \text{if case 1} ,\\
		0		& \text{if case 2},\\
		cr_A(1-P_D^{(r-1)}(A)-cr_A) & \text{otherwise}
    \end{cases}
  \label{eq:pSrA}
\end{equation}
The first two cases in Eqs. (\ref{eq:pDrA}) and \ref{eq:pSrA} indicate that client A, once finishing local training without crash, can always submit its update into the cache (for the upcoming aggregation). For case 3 (where $C<(1-R-C)(1-R)$ ), the chance for client A to be directly merged into cache equals to $(1-cr_A)\sigma_A^{(r-1)}$ because two conditions need to be satisfied: being undrafted/crashed last round and being picked this round. The situation that client A's entry in the bypass takes effect in round $r$ (i.e., $P_S^{(r)}(A)$ in case 3) only happens when the server ignores client A in both round $r-1$ and $r$ while A actually completed local training at round $r-1$.

For client B we have:
\begin{equation}
	P_D^{(r)}(B) = 
	\begin{cases}
		1 - cr_B 					& \text{if case 1} ,\\
		(1 - cr_B)(1-P_D^{(r-1)}(B))	& \text{if case 2},\\
		0						 	& \text{otherwise}
    \end{cases}
  \label{eq:pDrB}
\end{equation}

\begin{equation}
	P_S^{(r)}(B) = 
	\begin{cases}
		0 										& \text{if case 1} ,\\
		cr_B(1-P_D^{(r-1)}(B)-cr_B)				& \text{if case 2},\\
		1 - cr_B								& \text{otherwise}
    \end{cases}
  \label{eq:pSrB}
\end{equation}

The analysis for B is a bit more intuitive. In case 1(i.e., $C \geq 1-R$), client B cannot have any bypass entry available because it's update will always be merged into cache, and in case 2 client B 's entry in bypass takes effect only when it crash this round and was undrafted last round. In case 3, client B never gets picked (thus no direct update to cache) because too many undrafted clients are expected in each round, leaving no chance for B as it is the slowest in local training. But client still has a chance to contribute via the bypass, this happens in case it got training results in the previous round.

Considering the recurrence relation of $P_D^{(r)}(A)$ and $P_D^{(r)}(B)$ in Eqs. (\ref{eq:pDrA}) and (\ref{eq:pDrB}) in case 3 and 2 respectively, by resolving it we can derive the following expressions in terms of $r$:

\begin{equation}
  P_D^{(r)}(A) = \frac{(cr_A-1)^{n+1}+1-cr_A}{cr_A-2}, \textit{~for case 3}
  \label{eq:tx_PA}
\end{equation}

\begin{equation}
	P_D^{(r)}(B) = \frac{(cr_B-1)^{n+1}+1-cr_B}{cr_B-2}, \textit{~for case 2}
  \label{eq:tx_PB}
\end{equation}

Further, by defining $\sigma^{(k)}_A  = 1-P_D^{(k)}(A)$ and $\sigma^{(k)}_B = 1-P_D^{(k)}(B)$, we can reformulate $P_D^{(r)}(A)$, $P_S^{(r)}(A)$, $P_D^{(r)}(B)$ and $P_S^{(r)}(B)$ as:

\begin{equation}
	P_D^{(r)}(A) = 
	\begin{cases}
		1 - cr_A 					& \text{if case 1} ,\\
		1 - cr_A					& \text{if case 2},\\
		(1-cr_A)\sigma^{(r-1)}_A 	& \text{otherwise}
    \end{cases}
  \label{eq:pDrA_sigma}
\end{equation}

\begin{equation}
	P_S^{(r)}(A) = 
	\begin{cases}
		0 		& \text{if case 1} ,\\
		0		& \text{if case 2},\\
		cr_A(\sigma^{(r-1)}_A-cr_A) & \text{otherwise}
    \end{cases}
  \label{eq:pSrA_sigma}
\end{equation}

\begin{equation}
	P_D^{(r)}(B) = 
	\begin{cases}
		1 - cr_B 					& \text{if case 1} ,\\
		(1 - cr_B)\sigma^{(r-1)}_B	& \text{if case 2},\\
		0						 	& \text{otherwise}
    \end{cases}
  \label{eq:pDrB_sigma}
\end{equation}

\begin{equation}
	P_S^{(r)}(B) = 
	\begin{cases}
		0 										& \text{if case 1} ,\\
		cr_B(\sigma^{(r-1)}_B-cr_B)				& \text{if case 2},\\
		1 - cr_B								& \text{otherwise}
    \end{cases}
  \label{eq:pSrB_sigma}
\end{equation}

Combining all these results we can derive Eq. (\ref{eq:prA}) and Eq. (\ref{eq:prB}) in section \ref{sec_bias}.

\newpage
\section{Supplementary results of experiment}
\label{apdx:tabs}
\begin{table}[ht]

\centering
\caption{Best accuracy of the global model on Task 1}
\begin{tabular}{l l l l l l} 
 \hline
 \multicolumn{6}{c}{Best accuracy (Task 1: regression)}\\
 \multicolumn{6}{c}{Fully local}\\
 $cr$	& $C=0.1$	&$C=0.3$	&$C=0.5$ 	&$C=0.7$	&$C=1.0$\\ 
 \hline
 0.1	&0.6154 	&0.6308 	&0.5820 	&0.5423 	&0.5270 	\\
 0.3	&0.5806 	&0.6363 	&0.6145 	&0.5843 	&0.5443  	\\
 0.5	&0.5180 	&0.6043 	&0.6276 	&0.6181 	&0.5978 	\\
 0.7	&0.4443		&0.5480 	&0.6327 	&0.6409 	&0.6361 	\\
 \hline
 \multicolumn{6}{c}{FedAvg}\\
 $cr$	& $C=0.1$	&$C=0.3$	&$C=0.5$ 	&$C=0.7$	&$C=1.0$\\
 \hline
 0.1	&0.6055 	&0.6413 	&0.6411 	&0.6417 	&0.6424 	\\
 0.3	&0.6117 	&0.6418 	&0.6415 	&0.6419 	&0.6418  	\\
 0.5	&0.4432 	&0.6164 	&0.6421 	&0.6419 	&0.6413 	\\
 0.7	&0.3763 	&0.5576 	&0.6283 	&0.6413 	&0.6418 	\\
 \hline
 \multicolumn{6}{c}{FedCS}\\
 $cr$	& $C=0.1$	&$C=0.3$	&$C=0.5$ 	&$C=0.7$	&$C=1.0$\\
 \hline
 0.1	&0.6109 	&0.6415 	&0.6412 	&0.6417 	&0.6423 	\\
 0.3	&0.6077 	&0.6417 	&0.6416 	&0.6420 	&0.6418  	\\
 0.5	&0.4097 	&0.6073 	&0.6423		&0.6418 	&0.6413 	\\
 0.7	&0.2882 	&0.5999 	&0.6297 	&0.6293 	&0.6418 	\\
 \hline
 \multicolumn{6}{c}{SAFA}\\
 $cr$	& $C=0.1$	&$C=0.3$	&$C=0.5$ 	&$C=0.7$	&$C=1.0$	\\
 \hline
 0.1	&0.6419 	&0.6414 	&0.6413 	&0.6417 	&0.6423 	\\
 0.3	&0.6426 	&0.6419 	&0.6416 	&0.6417 	&0.6419  	\\
 0.5	&0.6423 	&0.6415 	&0.6422 	&0.6419 	&0.6415 	\\
 0.7	&0.6402 	&0.6422 	&0.6417 	&0.6412 	&0.6420 	\\
 \hline						
\end{tabular}
\label{tab:task1_acc}
\end{table}

\begin{table}[ht]
\centering
\caption{Synchronization Ratio and futility percentage on Task 1}
\begin{tabular}{l l l l l l} 
 \hline
 \multicolumn{6}{c}{SR / futility percentage (Task 1: regression)}\\
 \multicolumn{6}{c}{FedAvg}\\
 $cr$	& $C=0.1$	&$C=0.3$	&$C=0.5$ 	&$C=0.7$	&$C=1.0$	\\
 \hline
 0.1 	& 0.100/0.04 & 0.300/0.03 & 0.500/0.05 & 0.700/0.06 & 1.000/0.06 \\
 0.3 	& 0.100/0.09 & 0.300/0.14 & 0.500/0.14 & 0.700/0.17 & 1.000/0.16 \\
 0.5 	& 0.100/0.26 & 0.300/0.27 & 0.500/0.27 & 0.700/0.23 & 1.000/0.25 \\
 0.7 	& 0.100/0.33 & 0.300/0.36 & 0.500/0.31 & 0.700/0.38 & 1.000/0.35 \\
 \hline
 \multicolumn{6}{c}{FedCS}\\
 $cr$	& $C=0.1$	&$C=0.3$	&$C=0.5$ 	&$C=0.7$	&$C=1.0$	\\
 \hline
 0.1 	& 0.080/0.21 & 0.300/0.07 & 0.500/0.04 & 0.700/0.04 & 1.000/0.06 \\
 0.3 	& 0.100/0.17 & 0.300/0.12 & 0.500/0.14 & 0.648/0.32 & 1.000/0.16 \\
 0.5 	& 0.079/0.39 & 0.234/0.41 & 0.500/0.28 & 0.700/0.25 & 1.000/0.25 \\
 0.7 	& 0.078/0.49 & 0.300/0.33 & 0.500/0.35 & 0.644/0.51 & 1.000/0.35 \\
 \hline
 \multicolumn{6}{c}{SAFA}\\
 $cr$	& $C=0.1$	&$C=0.3$	&$C=0.5$ 	&$C=0.7$	&$C=1.0$	\\
 \hline
 0.1 	& 0.910/0.00 & 0.894/0.00 & 0.900/0.00 & 0.906/0.00 & 0.882/0.00 \\
 0.3 	& 0.722/0.00 & 0.714/0.00 & 0.716/0.01 & 0.678/0.00 & 0.700/0.00 \\
 0.5 	& 0.498/0.02 & 0.502/0.01 & 0.470/0.00 & 0.536/0.01 & 0.506/0.01 \\
 0.7 	& 0.368/0.04 & 0.346/0.02 & 0.362/0.03 & 0.342/0.04 & 0.354/0.03 \\
 \hline
\end{tabular}
\label{tab:task1_SR_futile}
\end{table}

\begin{table}[ht]
\centering
\caption{Best accuracy of the global model on Task 2}
\begin{tabular}{l l l l l l} 
 \hline
 \multicolumn{6}{c}{best accuracy (Task 2: CNN)}\\
 \multicolumn{6}{c}{Fully local}\\
 $cr$	& $C=0.1$	&$C=0.3$	&$C=0.5$ 	&$C=0.7$	&$C=1.0$	\\ 
 \hline
 0.1	&0.8849 	&0.9066 	&0.9026 	&0.9131 	&0.9019 	\\
 0.3	&0.8909 	&0.8932 	&0.8937 	&0.8909 	&0.9126  	\\
 0.5	&0.8649 	&0.8898 	&0.9021 	&0.8932 	&0.9081 	\\
 0.7	&0.8518 	&0.8956 	&0.9026 	&0.8959 	&0.9093 	\\
 \hline
 \multicolumn{6}{c}{FedAvg}\\
 $cr$	& $C=0.1$	&$C=0.3$	&$C=0.5$ 	&$C=0.7$	&$C=1.0$	\\
 \hline
 0.1	&0.9407 	&0.9664 	&0.9738 	&0.9766 	&0.9796 	\\
 0.3	&0.9326 	&0.9640 	&0.9705 	&0.9745 	&0.9755  	\\
 0.5	&0.9178 	&0.9532 	&0.9652 	&0.9696 	&0.9738 	\\
 0.7	&0.8818 	&0.9452 	&0.9534 	&0.9591 	&0.9672 	\\
 \hline
 \multicolumn{6}{c}{FedCS}\\
 $cr$	& $C=0.1$	&$C=0.3$	&$C=0.5$ 	&$C=0.7$	&$C=1.0$\\
 \hline
 0.1	&0.9423 	&0.9666 	&0.9732 	&0.9762 	&0.9791 	\\
 0.3	&0.9328 	&0.9626 	&0.9702 	&0.9741 	&0.9765  	\\
 0.5	&0.9232 	&0.9529 	&0.9650 	&0.9699 	&0.9321 	\\
 0.7	&0.8962 	&0.9434 	&0.9546 	&0.9599 	&0.9673 	\\
 \hline
 \multicolumn{6}{c}{SAFA}\\
 $cr$	& $C=0.1$	&$C=0.3$	&$C=0.5$ 	&$C=0.7$	&$C=1.0$	\\
 \hline
 0.1	&0.9748 	&0.9746 	&0.9764 	&0.9779 	&0.9787 	\\
 0.3	&0.9698 	&0.9696 	&0.9727 	&0.9753 	&0.9781  	\\
 0.5	&0.9658 	&0.9672 	&0.9686 	&0.9697 	&0.9714 	\\
 0.7	&0.9604 	&0.9632 	&0.9652 	&0.9603 	&0.9645 	\\
 \hline						
\end{tabular}
\label{tab:task2_acc}
\end{table}

\begin{table}[ht]
\centering
\caption{Synchronization Ratio and futility percentage on Task 2}
\begin{tabular}{l l l l l l} 
 \hline
 \multicolumn{6}{c}{SR / futility percentage (Task 2: CNN)}\\
 \multicolumn{6}{c}{FedAvg}\\
 $cr$	& $C=0.1$	&$C=0.3$	&$C=0.5$ 	&$C=0.7$	&$C=1.0$	\\
 \hline
 0.1 	& 0.100/0.03 & 0.300/0.04 & 0.500/0.05 & 0.700/0.05 & 1.000/0.05	\\
 0.3 	& 0.100/0.16 & 0.300/0.15 & 0.500/0.14 & 0.700/0.15 & 1.000/0.15 	\\
 0.5 	& 0.100/0.24 & 0.300/0.24 & 0.500/0.25 & 0.700/0.22 & 1.000/0.26 	\\
 0.7 	& 0.100/0.36 & 0.300/0.36 & 0.500/0.35 & 0.700/0.35 & 1.000/0.35	\\
 \hline
 \multicolumn{6}{c}{FedCS}\\
 $cr$	& $C=0.1$	&$C=0.3$	&$C=0.5$ 	&$C=0.7$	&$C=1.0$	\\
 \hline
 0.1 	& 0.092/0.04 & 0.300/0.05 & 0.500/0.06 & 0.693/0.06 & 1.00/00.05 	\\
 0.3 	& 0.099/0.16 & 0.296/0.16 & 0.500/0.15 & 0.692/0.15 & 1.000/0.16 	\\
 0.5 	& 0.100/0.23 & 0.297/0.28 & 0.500/0.25 & 0.700/0.25 & 0.990/0.25 	\\
 0.7 	& 0.100/0.33 & 0.300/0.33 & 0.495/0.35 & 0.700/0.36 & 0.990/0.36 	\\
 \hline
 \multicolumn{6}{c}{SAFA}\\
 $cr$	& $C=0.1$	&$C=0.3$	&$C=0.5$ 	&$C=0.7$	&$C=1.0$	\\
 \hline
 0.1 	& 0.896/0.00 & 0.902/0.00 & 0.891/0.00 & 0.900/0.00 & 0.894/0.00 	\\
 0.3 	& 0.704/0.00 & 0.710/0.00 & 0.704/0.00 & 0.707/0.00 & 0.709/0.00 	\\
 0.5 	& 0.524/0.01 & 0.517/0.01 & 0.521/0.01 & 0.521/0.01 & 0.509/0.01 	\\
 0.7 	& 0.341/0.04 & 0.351/0.04 & 0.359/0.04 & 0.342/0.04 & 0.350/0.04	\\
 \hline
\end{tabular}
\label{tab:task2_SR_futile}
\end{table}

\begin{table}[ht]
\centering
\caption{Best accuracy of the global model on Task 3}
\begin{tabular}{l l l l l l} 
 \hline
 \multicolumn{6}{c}{best accuracy (Task 3: SVM)}\\
 \multicolumn{6}{c}{Fully local}\\
 $cr$	& $C=0.1$	&$C=0.3$	&$C=0.5$ 	&$C=0.7$	&$C=1.0$	\\ 
 \hline
 0.1	&0.7793 	&0.6603 	&0.6307 	&0.6307 	&0.6307 	\\
 0.3	&0.7477 	&0.6363 	&0.6307 	&0.6307 	&0.6307  	\\
 0.5	&0.8419 	&0.6859 	&0.6339 	&0.6307 	&0.6307 	\\
 0.7	&0.9530 	&0.7886 	&0.6673 	&0.6491 	&0.6442 	\\
 \hline
 \multicolumn{6}{c}{FedAvg}\\
 $cr$	& $C=0.1$	&$C=0.3$	&$C=0.5$ 	&$C=0.7$	&$C=1.0$	\\
 \hline
 0.1	&0.9935 	&0.9942 	&0.9962 	&0.9992 	&0.9992 	\\
 0.3	&0.9961 	&0.9961 	&0.9962 	&0.9963 	&0.9992  	\\
 0.5	&0.9961 	&0.9960 	&0.9961 	&0.9962 	&0.9963 	\\
 0.7	&0.9961 	&0.9961 	&0.9961 	&0.9957 	&0.9962 	\\
 \hline
 \multicolumn{6}{c}{FedCS}\\
 $cr$	& $C=0.1$	&$C=0.3$	&$C=0.5$ 	&$C=0.7$	&$C=1.0$\\
 \hline
 0.1	&0.9959 	&0.9962 	&0.9961 	&0.9991 	&0.9993 	\\
 0.3	&0.9961 	&0.9961 	&0.9962 	&0.9963 	&0.9992  	\\
 0.5	&0.9961 	&0.9962 	&0.9959 	&0.9962 	&0.9962 	\\
 0.7	&0.9961 	&0.9776 	&0.9960 	&0.9960 	&0.9960 	\\
 \hline
 \multicolumn{6}{c}{SAFA}\\
 $cr$	& $C=0.1$	&$C=0.3$	&$C=0.5$ 	&$C=0.7$	&$C=1.0$	\\
 \hline
 0.1	&0.9962 	&0.9961 	&0.9962 	&0.9992 	&0.9992 	\\
 0.3	&0.9960 	&0.9961 	&0.9962 	&0.9991 	&0.9991 	\\
 0.5	&0.9959 	&0.9961 	&0.9962 	&0.9961 	&0.9962 	\\
 0.7	&0.9960 	&0.9934 	&0.9961 	&0.9958 	&0.9960 	\\
 \hline						
\end{tabular}
\label{tab:task3_acc}
\end{table}

\begin{table}[ht]
\centering
\caption{Synchronization Ratio and futility percentage on Task 3}
\begin{tabular}{l l l l l l} 
 \hline
 \multicolumn{6}{c}{SR / futility percentage (Task 3: SVM)}\\
 \multicolumn{6}{c}{FedAvg}\\
 $cr$	& $C=0.1$	&$C=0.3$	&$C=0.5$ 	&$C=0.7$	&$C=1.0$	\\
 \hline
 0.1 	& 0.100/0.05 & 0.300/0.05 & 0.500/0.05 & 0.700/0.05 & 1.000/0.05 \\
 0.3 	& 0.100/0.15 & 0.300/0.15 & 0.500/0.15 & 0.700/0.15 & 1.000/0.15 \\
 0.5 	& 0.100/0.25 & 0.300/0.25 & 0.500/0.25 & 0.700/0.25 & 1.000/0.25 \\
 0.7 	& 0.100/0.35 & 0.300/0.35 & 0.500/0.35 & 0.700/0.35 & 1.000/0.35\\
 \hline
 \multicolumn{6}{c}{FedCS}\\
 $cr$	& $C=0.1$	&$C=0.3$	&$C=0.5$ 	&$C=0.7$	&$C=1.0$	\\
 \hline
 0.1 	& 0.100/0.05 & 0.299/0.05 & 0.499/0.05 & 0.697/0.05 & 0.998/0.05 \\
 0.3 	& 0.099/0.15 & 0.300/0.15 & 0.499/0.15 & 0.698/0.15 & 0.998/0.15 \\
 0.5 	& 0.099/0.24 & 0.300/0.25 & 0.499/0.25 & 0.697/0.25 & 0.998/0.25 \\
 0.7 	& 0.100/0.36 & 0.300/0.35 & 0.498/0.36 & 0.699/0.36 & 0.996/0.36\\
 \hline
 \multicolumn{6}{c}{SAFA}\\
 $cr$	& $C=0.1$	&$C=0.3$	&$C=0.5$ 	&$C=0.7$	&$C=1.0$	\\
 \hline
 0.1 	& 0.901/0.00 & 0.900/0.00 & 0.900/0.00 & 0.901/0.00 & 0.901/0.00 	\\
 0.3 	& 0.703/0.00 & 0.700/0.00 & 0.701/0.00 & 0.702/0.00 & 0.703/0.00 	\\
 0.5 	& 0.512/0.01 & 0.514/0.01 & 0.516/0.01 & 0.513/0.01 & 0.512/0.01 	\\
 0.7 	& 0.345/0.04 & 0.343/0.04 & 0.344/0.04 & 0.342/0.04 & 0.344/0.04	\\
 \hline
\end{tabular}
\label{tab:task3_SR_futile}
\end{table}

\end{document}